\documentclass[a4paper]{aa}
\usepackage{graphicx}
\usepackage[]{natbib}
\bibpunct{(}{)}{;}{a}{}{,}

\def \Ha {H$\alpha$~}
\def \Mdot {$\dot M$\,}

\def \Yhe {$Y_{\rm He}$\,} 
\def \kms {km~s$^{\rm -1}$\,}
\def \he {He{\sc ii} $\lambda$6560\,}
\def \hee {He{\sc ii} $\lambda$6527\,} 
\def \Rstar {$R_\star$\,}
\def \Teff {$T_{\rm eff}$\,}
\def \vinf {$v_\infty$\,}
\def \vsini {$v \sin i$\,}
\def \ve    {$v_{\rm e}$}
\def \Msun {$M_\odot$\,}
\def \logg {$\log g$\,}
\def \vsys {$v_{\rm sys}$\,}

\def \logl {$\log L/L_\odot$\,}

\def 
\Mdu {\cdot 10^{-6} {\rm M_{\odot}/yr}} 
\def \Rstare {R_\star}

\def \vinfe {v_\infty}
\def \Mdote {\dot M}
\def \We    {$\overline{W_{\rm em}}$}
\def \beq{\begin{equation}}
\def \eeq{\end{equation}}
\def \ben{\begin{enumerate}} 
\def \een{\end{enumerate}} 
\begin{document}
\title{Bright OB  stars in the Galaxy}
\subtitle{II. Wind variability in O supergiants as traced by \Ha}

\author{N. Markova\inst{1}, 
           J. Puls\inst{2}, 
	   S. Scuderi\inst{3}, 
	   \and H. Markov\inst{1}}
\offprints{N. Markova,\\ \email{rozhen@mbox.digsys.bg}}

\institute{Institute of Astronomy, 
        Bulgarian National Astronomical Observatory, 
	P.O. Box 136, 4700 Smoljan, Bulgaria\\ 
	\email{nmarkova@libra.astro.bas.bg}
	\and Universit\"{a}ts-Sternwarte,
	      Scheinerstrasse 1, D-81679 M\"unchen, Germany\\
	\email{uh101aw@usm.uni-muenchen.de}
	\and Osservatorio Astrofisico di Catania, Viale A. Doria 6,
	I-95125, Catania, Italy\\ 
	\email{scuderi@ct.astro.it}}
\date{Received; Accepted }

\abstract {In this study we investigate the line-profile variability (lpv)
of \Ha for a large sample of O-type supergiants (15 objects between O4 and
O9.7), in an objective, statistically rigorous manner. We employed the
Temporal Variance Spectrum (TVS) analysis, developed by \citet{Fullerton96}
for the case of photospheric absorption lines and modified by us to take
into account the effects of wind emission.  By means of a comparative
analysis we were able to put a number of constraints on the properties of
this variability -- quantified in terms of a mean and a newly defined
fractional amplitude of deviations -- as a function of stellar and wind
parameters. The results of our analysis show that all the stars in the
sample show evidence of {\it significant} lpv in \Ha, mostly dominated by
processes in the wind. The variations occur between zero and 0.3~\vinf
(i.e., below $\sim1.5$~\Rstar), in good agreement with the results from
similar studies.\\ 
A comparison between the observations and corresponding line-profile
simulations indicates that for stars with {\it intermediate} wind densities
the properties of the \Ha variability can be explained by simple models,
consisting of coherent or broken shells (blobs) uniformly distributed over
the wind volume, with an intrinsic scatter in the maximum density contrast
of about a factor of two. For stars at lower and higher wind densities, on
the other hand, we found certain inconsistencies between the observations
and our predictions, most importantly concerning the mean amplitude and the
symmetry properties of the TVS. This disagreement might be explained with
the presence of coherent large-scale structures (e.g., CIRs), partly
confined in a volume close to the star.\\ 
Interpreted in terms of a variable mass-loss rate, the observed variations of
\Ha indicate changes of $\pm$4\% with respect to the mean value of \Mdot for
stars with stronger winds and of $\pm$ 16\% for stars with weaker winds. The
effect of these variations on the corresponding wind momenta is rather
insignificant (less than 0.16~dex), increasing only the local scatter
without affecting the main concept of the Wind Momentum Luminosity
Relationship.  

\keywords{stars:
early type -- stars: mass loss -- stars: winds, outflows -- stars: activity
-- methods: data analysis} } 

\authorrunning{Markova et al.}

\titlerunning{Wind variability in O stars } 

\maketitle

\section{Introduction}

The basic philosophy underlying present day hot star model atmospheres 
contains the assumption of a globally stationary and spherically symmetric
stellar wind with a smooth density stratification. Although these models are
generally quite successful in describing the overall wind properties, there
are theoretical considerations, supported by numerous observational 
evidences, which indicate that hot stars winds are very far from being 
smooth and stationary.

The most common approach used to study wind variability in optical and UV
domains is to follow line-profile variability (lpv) of one or several
spectral lines, formed in different regions, in order to determine relevant
time-scales and variability patterns and thus to obtain some insight into
the nature and the physical origin of the variations. This kind of surveys
requires long sets of stellar spectra with high S/N ratio and high temporal 
resolution, which implies that only few objects have been investigated so
far. Through such investigations clear evidence for the presence of
large-scale time-dependent wind perturbations (e.g., in the form of Discrete
Absorption Components, DACs) was found in UV
\citep{Prinja92,Massa95,Prinja96,Kaper99,Prinja02} and optical
\citep{Fullerton92,Prinja94, Rauw01,Prinja01,Markova02} spectra of many O
and early B stars. Since DACs have been observed in WR stars \citep{PS92}
and in an LBV \citep{Markova86} as well, they are thought to be a
fundamental property of radiative driven stellar winds.

Another source of lpv in hot stars winds are small-scale structures (clumps)
which are believed to result from strong instabilities in the wind itself 
\citep{O88, Feldm95, OP99}. While the clumped nature of WR winds was
unambiguously proven by observations, the presence of clumps in O-star winds
has so far relied on indirect evidence only
\citep{Crowther02,BG02,Markova04}. 

Wind structures and temporal variability are among the most important
physical processes that may significantly modify the mass loss rates derived
from observations. Since accurate mass loss rates are crucial for
evolutionary studies (e.g., \citealt{Meynet94}) and for extra-galactic distance
determinations (via the Wind Momentum Luminosity Relationship 
(WLR), cf. \citealt{KP2000}), it is particularly important to know to what
extent the outcome of these studies might be influenced by uncertainties in
\Mdot due to the effects of wind structures and variability. Indeed,
\citet{Kudritzki99} has noted that wind variability is not expected to
affect the concept of the WLR significantly. However, this
suggestion is based on results obtained via a detailed investigation of one
object alone, while similar data for a large number of stars of different
spectral types and luminosity classes are needed to resolve the problem
adequately.

Following the outlined reasoning, a project to study the effects of wind
structure and variability in Galactic O-type stars has been recently started
by our group. While in a previous paper \citep[Paper~I]{Markova04} we have
dealt with problems concerning the WLR and the effects of wind clumping, in
the present one we address the question of wind variability as traced by \Ha
and the dependence (if any) of the properties of this variability on
fundamental stellar and wind parameters. In particular, in Sect.~\ref{obs}
we describe the observational material and its reduction. In
Sect.~\ref{method} we outline the method used to detect and analyze the \Ha
lpv. In Sect.~\ref{Ha_time}, \ref{Ha_sp} and \ref{wlr} the results of our
analysis are presented in detail while in Sect.~\ref{simul} the outcomes of
some simple 1- and 2-D simulations are described.  In Sect.~\ref{summary} we
summarize the major results and give some comments and conclusions.  

\section{Observations and data reduction}
\label{obs}

Our sample consists of 15 Galactic supergiants with spectral classes from O4
to O9.7, all drawn from the list of stars analyzed by \citet{Markova04} in 
terms of their mass-loss and wind momentum rates. Table~\ref{log} lists the
objects along with some of their stellar and wind parameters, as used in the
present study. All data are taken from Paper~I.
\begin{table*}
\caption{Stellar and wind parameters of the sample stars used in the present
study. All data are taken from \citet{Markova04}.}
\label{log}
\tabcolsep1.5mm
\begin{tabular}{llrlrllcrcc}
\hline 
\hline
~\\
\multicolumn{1}{c}{Object}
&\multicolumn{1}{l}{Sp}
&\multicolumn{1}{l}{\vsys}
&\multicolumn{1}{c}{\Teff}
&\multicolumn{1}{c}{\Rstar}
&\multicolumn{1}{c}{\logg}
&\multicolumn{1}{c}{\Yhe}
&\multicolumn{1}{c}{$\log L$}
&\multicolumn{1}{c}{\vsini}
&\multicolumn{1}{c}{\vinf}
&\multicolumn{1}{c}{$\beta$}
\\
\hline
\\
HD 190\,429A  &O4If+     &-36 &39\,200 &20.8  &3.65  &0.14 &5.97 &135  &2\,400 &0.95 \\
			  			   	        
HD 16\,691    &O4If      &-51 &39\,200 &19.8  &3.65  &0.10 &5.92 &140  &2\,300 &0.96 \\
			  			   	        
HD 14\,947    &O5If      &-56 &37\,700 &25.6  &3.56  &0.20 &6.08 &133  &2\,300 &0.98 \\
			  			   	        
HD 210\,839   &O6If      &-71 &36\,200 &23.0  &3.48  &0.10 &5.91 &214  &2\,200 &1.00 \\
			  			   	        
HD 192\,639   &O7Ib(f)   &-7  &34\,700 &17.2   &3.39 &0.20 &5.59 &110  &2\,150 &1.09 \\
							        
HD 17\,603    &O7.5Ib(f) &-40 &34\,000 &25.2   &3.35 &0.12 &5.88 &110  &1\,900 &1.05\\
							        
HD 24\,912    &O7.5I(f)  &59  &34\,000 &25.2   &3.35 &0.15 &5.88 &204  &2\,400 &0.78 \\
							        
HD 225\,160   &O8Ib(f)   &-40 &33\,000 &22.4   &3.31 &0.12 &5.73 &125  &1\,600 &0.85 \\
			  	 			        
HD 338\,926   &O8.5Ib    &-9  &32\,500 &22.7   &3.27 &0.12 &5.72 & 80  &2\,000 &1.00\\
							        
HD 210\,809   &O9Iab     &-90 &31\,700 &19.6   &3.23 &0.14 &5.54 &100  &2\,100 &0.91 \\

HD 188\,209   &O9.5Iab   &-16 &31\,000 &19.6   &3.19 &0.12 &5.51 & 87  &1\,650 &0.90 \\
							        
BD+56 739     &O9.5Ib    &-5  &31\,000 &19.6   &3.19 &0.12 &5.51 & 80  &2\,000 &0.85 \\
							        
HD 209\,975   &O9.5Ib    &-18 &31\,000 &19.2   &3.19 &0.10 &5.49 & 90  &2\,050 &0.80 \\
							        
HD 218\,915   &O9.5Iab   &-84 &31\,000 &19.6   &3.19 &0.12 &5.51 & 80  &2\,000 &0.95\\
			  			   	        
HD 18\,409    &O9.7Ib    &-51 &30\,600 &15.7   &3.17 &0.14 &5.29 &110  &1\,750 &0.70 \\
\hline
\end{tabular}
\end{table*}

A total of 82 high-quality \Ha spectra ($R=15\,000$) of the sample stars
were collected between 1997 and 1999. The observations were obtained at the
Coud\'e focus of the 2m RCC telescope at the National Astronomical
Observatory (Bulgaria) using an ELECTRON CCD
(520$\times$580,22$\times$24$\mu$) and a PHOTOMETRIC CCD
(1024$\times$1024,24$\mu$).\footnote{The use of different detectors is not
expected to bias the homogeneity of our sample because the noise
characteristics of these two devices are practically the same. The
root-mean-square ($rms$) read-out noise of the ELECTRON CCD is 3 electrons
per pixel (i.e 1.5 ADU with 2 electrons per ADU) while the $rms$ read-out
noise of the PHOTOMETRIC CCD is 3.3 electrons per pixel (2.7 ADU with 1.21
electrons per ADU).} For all stars but one the S/N ratio, averaged within
each spectral time series, lies between 150 to 250, while in the case of 
HD~190429 it is $\sim$100.  

The temporal sampling of the data for each target is not systematic but
random, with typical values of the minimum and maximum time intervals
between successive spectra of 1 to 2 and 7 to 8 months, respectively. In
several cases observations with a time-resolution of 1 to 5 days are also
available, but in none of these cases these observations dominate the 
corresponding time series. Thus, we expect the results of our survey to be
sensitive to variations which occur on a time-scale which is significantly
larger than the corresponding wind flow time (of the order of a couple of
hours).

More information about the observational material  and  its reduction
can be found in \citet{MV} and in Paper~I. In particular, to reduce the
observations we have followed a standard procedure (developed in IDL) 
which includes: bias subtraction, flat-fielding, cosmic ray hits removal, 
wavelength calibration, correction for heliocentric radial velocity, water 
vapor lines removal and re-binning to a step size of 0.2~\AA\, per pixel.
   
\section{Methodology and measurements} 
\label{method}

Since we were going to study a large number of objects and since in many
cases our observations were not systematic but with large temporal gaps in 
between, from the onset of this investigation on we recognized that our
ability to characterize the wind variability of individual targets would be
restricted, e.g., we would not be able to determine time-scales and
variability patterns. Moreover, to work effectively, we would need to employ
some simple and fast method both to detect and quantify lpv and to constrain
the properties of this variability as a function of fundamental stellar and
wind parameters of the sample stars. 

\citet{Fullerton96} have developed a very sensitive and rigorous method to
handle lpv that is poorly sampled in time. This method, called Temporal
Variance Spectrum (TVS) analysis, was proven to be a powerful tool to detect
and compare lpv of pure absorption profiles. However, its potential with
respect to profiles influenced by wind emission has not been tested
systematically so far.  Although the application of the TVS technique to
detect lpv in emission lines does not seem to pose serious problems (e.g.,
\citealt{Kaufer96,Kaper97,Markova02}), its implication for the objectives of
a comparative analysis (e.g. to compare the strength of lpv in different
lines or different stars) certainly needs to be carefully investigated.

To study the \Ha variability of the stars in our sample we modified the main
philosophy of the TVS analysis in order to take into account the effect of 
wind emission.

To compute the $TVS$ of \Ha as a function of velocity across the line and to
determine the velocity width over which {\it significant} variability
occurs, $\Delta V$, we followed \citet{Fullerton96} but assumed that the
noise is dominated by photon noise.\footnote{This assumption seems to be
justified because we rely on Coud\'e spectra of relatively high quality.} In
this case, the $TVS$ for the pixels in column $j$ (i.e., at
wavelength/velocity $j$) is calculated from
\beq
TVS_{j} = \sum_{i}^N \frac {w(i)(S_{ij}  - \overline{S_{j}})^{2}}
{S_{ij}(N - 1)} 
\eeq
where  $\overline{S_{j}}$ is the weighted mean spectrum for the $j$-th
pixel, averaged over a time series of $N$ spectra, and given by
\beq
\overline{S_{j}} = \frac{\sum_{i}^N S_{ij} w_{i}}{N}.
\eeq
with the weighting factors, $w_{i}$, given by
\beq
w_{i} = \bigl(\frac{\sigma_0}{\sigma_{ic}}\bigr)^{2}
\eeq
where
\beq
\sigma_0 = \left(\frac{1}{N} \sum_{i}^N{\sigma_{ic}^{-2}}\right)^{-1}
\eeq
and $\sigma_{ic}$  is the value of the noise in the $i$-th 
spectrum, averaged over a certain number of continuum pixels (40 in our case). 

The $RMS$ deviations ($RMS = TVS^{0.5}$) as a function of velocity across 
\Ha for the time series of each target are shown in the top panels of
Figures~\ref{hots} to \ref{lts}. The level of deviations in the continuum,
$\sigma_0$, is represented by a dashed line, while the threshold of {\it
significant} lpv, fixed at the corresponding 99\% confidence level of the
$\sigma_0^{2}\chi_{N-1}^{2}$ distribution, is 
marked with a dashed-dotted line. We want to stress here that although our 
implementation is in terms of $TVS^{0.5}$, hereafter we shall continue to
refer to the ``TVS'' and the ``TVS analysis'', respectively.\footnote{Root
mean square deviations have been used instead of the TVS itself since the
former quantity scales linearly with the size of the deviations. Thus, it is
more appropriate for a direct comparison of the strength of lpv in various
stars (see also Fullerton et al.).} 

To localize the \Ha lpv in velocity space we used the ``blue'' and ``red''
velocity limits of {\it significant} variability, $v_{\rm b}$ and $v_{\rm
r}$, introduced by \citet{Fullerton96}. The measurements have been performed
interactively to fix the positions of the two points where the TVS crosses
the horizontal line representing the threshold of {\it significant} lpv. The
accuracy of these measurements depends on the quality of the data used and
on the strength of lpv. For example, in the limiting case of a strong lpv
(i.e., a TVS with large amplitudes and steep spectral gradients) the
accuracy of the individual measurements might be as good as $\pm$20 \kms.
Alternatively, in the case of a weak lpv (e.g., with amplitudes just above
the threshold of {\it significant} variability) the determination of the
velocity limits might become so uncertain that different positions of almost
similar probability may exist for each limit. In these latter cases and in
order to assess the effects of such uncertainties on the outcomes of our
analysis, we provide two couples of estimates for $v_{\rm b}$ and $v_{\rm
r}$. These two sets of values, expressed in \kms, are listed in Column~6 of
Table~\ref{lpv} as a first and a second entry.  We consider the first entry
as the more reliable one and will refer to it as the ``conservative case''.

As a by-product of the measurement of $v_{\rm b}$ and $v_{\rm r}$, we obtain
the total velocity width over which {\it significant} variability in \Ha
occurs, $\Delta V = (v_{\rm r} - v_{\rm b}$). In order to find some
constraints on the distribution of the lpv in {\it physical} space, we
furthermore determined the radial distance $r_{\rm max}$, where
$v(r=r_{\rm max}) = v_{\rm b}$, assuming that the wind velocity obeys a
standard law of the form
\beq
v(r) = \vinfe \bigl( 1 - b \frac{\Rstare}{r} \bigr)^{\beta},
\eeq
\begin{equation} 
b = 1 - \bigl(\frac{v_{min}}{\vinfe} \bigr)^{1/\beta},
\label{vellaw}
\eeq
with $\beta$ and \vinf from Table~\ref{log} and $v_{\rm min}$ = 1.0 \kms.  
The obtained estimates of $r_{\rm max}$, expressed in units of \Rstar, 
are given in Column~7 of Table~\ref{lpv}. We are aware of the fact that 
emission variability is difficult to localize and could in principle be due
to the net effect of fluctuations that occur in different locations under
different conditions and therefore consider these estimates as upper limits
only.

To quantify and compare lpv, \citet{Fullerton96} have introduced two 
parameters, called mean and fractional amplitude of deviations, $A_{\rm
lpv}$ and $a_{\rm lpv}$. In the following we will refer to these
quantities as to $A_{F}$ and $a_{\rm F}$ (with ``F'' referring to Fullerton).
The first parameter is expressed in units of the normalized continuum flux, 
while the second one is a dimensionless quantity. The authors define these 
quantities as follows: 
\beq
A_{F} = \frac{1}{\Delta V} \int_{v_b}^{v_r} TVS_{j}^{0.5}\,dv
\label{A_lpv}
\eeq
\beq
a_{F} = \frac{100\int_{v_b}^{v_r} \left( TVS_{j} - \sigma_0^2\right)^{0.5}\,dv}
{\int_{v_b}^{v_r} \left|\overline{S}_{j} -1\right|\,dv}
\label{a_lpvF}
\eeq
The above expressions imply the following. {\bf First}, the mean amplitude
of deviations is independent of profile type and thus can be used for
evaluating the statistical significance of lpv both in absorption and in
emission profiles. {\bf Second}, the fractional amplitude depends (via the
denominator) on the strength of the underlying spectral feature but does not
make any difference between profiles in absorption and in emission.
Particularly, it becomes a non-monotonic function of wind strength, with a
maximum in those regions where the wind-emission has (more or less)
completely filled in the photospheric absorption, i.e., where the net
equivalent width within $[v_b, v_r]$ is close to zero. Therefore, this
quantity is inappropriate for investigating profiles which are influenced by
wind emission of different extent. Actually, this problem has already been
outlined by Fullerton et al.

In order to optimize the fractional amplitude to account for the
systematic difference in the strength of \Ha as a function of wind strength,
we decided to normalize the integral over the TVS to a quantity which we
called ``Fractional Emission Equivalent Width'' (FEEW).  With this new
definition of the fractional amplitude, now denoted by $a_{\rm N}$ to
distinguish it from Fullerton's parameter $a_{\rm F}$, this quantity is a
measure of the {\it observed degree of variability per unit fractional wind
emission}. ``Fractional'' refers here to the observed range of significant
variability, $[v_b,v_r]$. Formally, $a_{\rm N}$ is given
by\footnote{In this expression, we have accounted only for 
uncertainties caused by photon noise while the error due to small
differences in the continuum level of individual spectra in a given time
series is neglected.}
\beq
a_{N} = \frac{100\int_{v_b}^{v_r} \left( TVS_{j} - 
\sigma_0^2\right)^{0.5}\,dv}
{\int_{v_b}^{v_r} \left(\overline{S}_{j} - 1\right)\,dv - 
\int_{v_b}^{v_r}(S_{j}^{\rm phot}-1)\,dv}
\label{a_lpvN}
\eeq
The first term in the denominator of Eq.\ref{a_lpvN} represents the
fractional equivalent width of the observed profile (positive for
emission and negative for absorption), while the second one gives the
fractional equivalent width of the photospheric component of \Ha (always
negative). In total, the denominator thus gives the fractional {\it wind}
emission (always positive). A further discussion of $a_{\rm N}$ is given in
Sect.~\ref{simul}.

The mean amplitude, as defined by Eq.~\ref{A_lpv}, on the other hand, does
not seem to pose any problem concerning an assessement of the statistical
significance of variability across H$\alpha$. In their original study,
Fullerton et al. have noted that this quantity cannot serve as a comparative
tool because it does not account for differences in the strength of the
underlaying absorption feature. In contrast, in the case of \Ha from O-type
supergiants the mean amplitude might depend on the wind
strength\footnote{E.g., the numerator in Eq.~\ref{A_lpv} is expected to
react on wind density (the higher the density the larger the emitting
volume).} and might therefore become of interest as well, in order to
examine and to compare the wind variability in stars of various spectral
types.  Motivated by this possibility we re-defined the mean amplitude to
account (partially) for differences in the overall quality (i.e., in S/N) of
the time series of the sample stars, by subtracting $\sigma_0^2$ from the
TVS,
\beq
A_{N} = \frac{100}{\Delta V} \int_{v_b}^{v_r} \left( TVS_{j} -\sigma_0^{2} \right)^{0.5}\,dv.
\label{A_N}
\eeq

\medskip
\noindent
The photospheric profiles of \Ha required to derive the values of $a_{\rm
N}$ have been selected from a grid of plane-parallel models in dependence of
the particular stellar parameters (Table~1, see also Paper~I). Note that the
(relative) uncertainty of the denominator becomes rather large in those
cases where the wind-emission is only marginal, since in this case the
errors introduced by uncertainties in the stellar parameters (affecting the
actual choice of the photospheric profiles) become significant. 

To estimate the uncertainty in $a_{\rm N}$ we followed \citet{Fullerton96}
but used a re-formulation (derived by A.~Fullerton, priv. com.) of their Eq.
16.  Additionally, we assumed that the errors in both $\sigma_0$ and FEEW
are negligible and that the accuracy of the deviations for each pixel $j$
within \Ha is identical and equals $\sigma_0$.\footnote{The latter
assumption is justified since we rely on Coud\'e spectra of relatively high
signal to noise ratio.} Under these circumstances, standard error
propagation gives:
\begin{eqnarray}
\sigma(a_{N}) & = &\sigma_0 \frac{100\, \Delta v}{FEEW}      
\left[ \frac{1}{2(N - 1)} \right]^{0.5} \times \nonumber \\
&\times& \left[ \sum_{j}^n \frac{1}{\frac{TVS_j}{\sigma_0^2} - 1} \right]^{0.5} 
\label{err}
\end{eqnarray}
where $j$ runs over all the pixels between $v_{\rm b}$ and $v_{\rm r}$,
while $N$ and $\Delta v$ denote the number of spectra in the time series and
the discretized integration step, respectively. In our case $\Delta v$
$\sim9$ \kms (not to be confused with the total velocity width $\Delta V$!).

The factor of 100 appearing in Eqs.~\ref{a_lpvF} to \ref{err} converts the
corresponding quantities to a percentage. The estimates of $\sigma_0$, $A_{N}$
and $a_{\rm N}\pm\sigma(a_{\rm N})$) for each sample star are listed in 
Table~\ref{lpv}, Columns~5, 13 and 14, respectively. 

To assess the contribution of changes in \Ha line strength to the lpv
detected by the TVS analysis, we estimated the mean value of the net wind
emission, \We, by subtracting the equivalent width (EW) of the photospheric
profile, $W_{\rm phot}$, from the EW of the time-averaged observed profiles.

The observed equivalent widths were measured by integrating the line flux
between limits which were set interactively, judging by eye the
extension of the emission/absorption wings. These limits did not change 
for a given star but could vary for different stars. The internal precision 
of individual EW measurements, estimated in the way described in \citet{MV}, 
is better than 10\%. The EW of the photospheric component was calculated
by integrating over the appropriate synthetic profile. The
\We\, estimates and their (standard) error are given in Column~8 of
Tab.~\ref{lpv}, together with the EW of the photospheric components.

Since in O-type supergiants \Ha originates from processes taking place in
the wind {\it and} in the photosphere, contributions from {\it absorption}
lpv to the observed lpv might be expected (via the photospheric components
of \Ha and \he). To investigate this possibility we consulted the literature
(particular references are given below) concerning the presence of
absorption lpv in our sample.  In addition and as a secondary criterion, we
used the TVS of the \hee absorption line located at about 1650 \kms
blue-wards of H$\alpha$. In those cases where the results of our TVS
analysis of \hee did not agree with the results from the literature, the
latter were adopted.\footnote{Such inconsistencies may occur because the
temporal sampling of our observations is not well-suited for studying lpv on
short time scales (e.g., hours) which seem to be typical for absorption lpv
in O-type stars \citep{Fullerton96}.} 

To obtain constraints on the variability of mass-loss rates, for each star
we determined lower and upper limits for \Mdot. This has been done by
fitting those \Ha profiles which display the smallest and the largest wind
emission present in the given time-series, by means of synthetic profiles.
These have been calculated using stellar and wind parameters from
Table~\ref{log} and employing the same method as used in Paper~I. The
accuracy of the \Mdot determinations equals $\pm$20\% for stars with \Ha in
emission and $\pm$30\% for stars with \Ha in absorption \citep{Markova04}.
The estimates of \Mdot$_{\rm min}$ and \Mdot$_{\rm max}$ as well as the
amplitude of the \Mdot variability (given in percent of \Mdot$_{\rm min}$)
are listed in Columns 9 to 11 of Table~\ref{lpv}.

To quantify the wind strength we finally calculated the ``mean wind
density'', $<\rho>$, using data given in Table~\ref{log} and
Table~\ref{lpv} (Column~12), by means of
\beq
<\rho> = \frac{\Mdote}{4 \pi (1.4 \Rstare)^2 \vinfe},
\label{rhobar}
\eeq
i.e., we considered the density at a typical location of 1.4~\Rstar. 

\section{TVS analysis of \Ha line-profile variability}
\label{Ha_time}
\subsection{Stars of early spectral types} 

Within our sample, the O4/5 supergiants HD~190\,429A, HD~16\,691 and
HD~14\,947 constitute the group with the highest effective temperature. Their
\Ha profiles, displayed in Fig~\ref{hots}, appear to be completely in emission 
and consist of a well-developed emission core superimposed onto relatively
strong and extended emission wings. The profiles are slightly asymmetric
with a red wing being steeper than the blue one and a peak emission
red-shifted with respect to the stellar rest frame.

For all stars, the mean amplitudes of deviations are significant at the 99\%
confidence level, indicating genuine lpv in H$\alpha$. Whereas for
HD~190\,429A and HD~16\,691 the variations in \Ha are distributed
preferentially on the blue side of the emission peak, for HD~14\,947 they
extend almost symmetrically with respect to it. For HD~190\,429A two values
for $v_{\rm r}$ are provided, because of the rather small amplitudes of
deviations at the red edge of the TVS.

Since the TVS analysis does not show evidence for {\it significant} lpv in
\hee (see upper panels of Fig.~\ref{hots}), we suggest that the variations 
observed in \Ha are mostly (if not completely) due to processes in the wind.
In the particular case of HD~190\,429A, this assumption is supported by 
results reported by \citet{Fullerton96}, indicating that the photospheric 
C{\sc iv}$\lambda$5801 line of this star does not show signatures of
noticeable lpv. 

Our measurements indicate that in each of the stars considered here the 
observed lpv in \Ha has been accompanied by real (i.e., extending the
measurement errors of 10\%) variations in the equivalent width. In terms of
the adopted model these variations in the \Ha line strength can be
reproduced by variations in \Mdot ranging from 8\% (HD~190429A) to 10\%
(HD~14947).

\begin{figure*}
\begin{minipage}{5.9cm}
\resizebox{\hsize}{!}
{\includegraphics{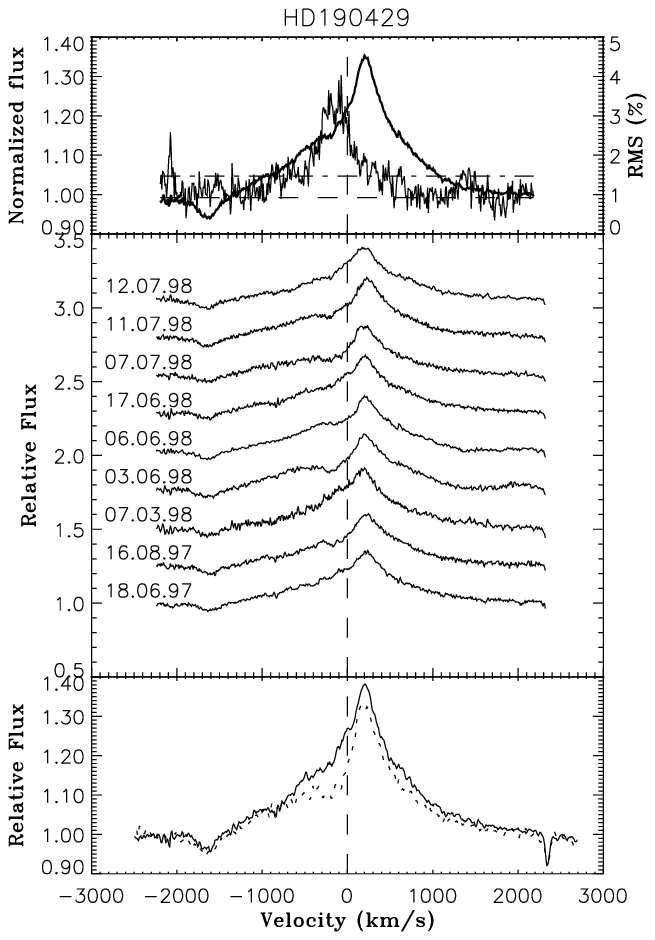}}
\end{minipage}
\hfill
\begin{minipage}{5.9cm}
\resizebox{\hsize}{!}
{\includegraphics{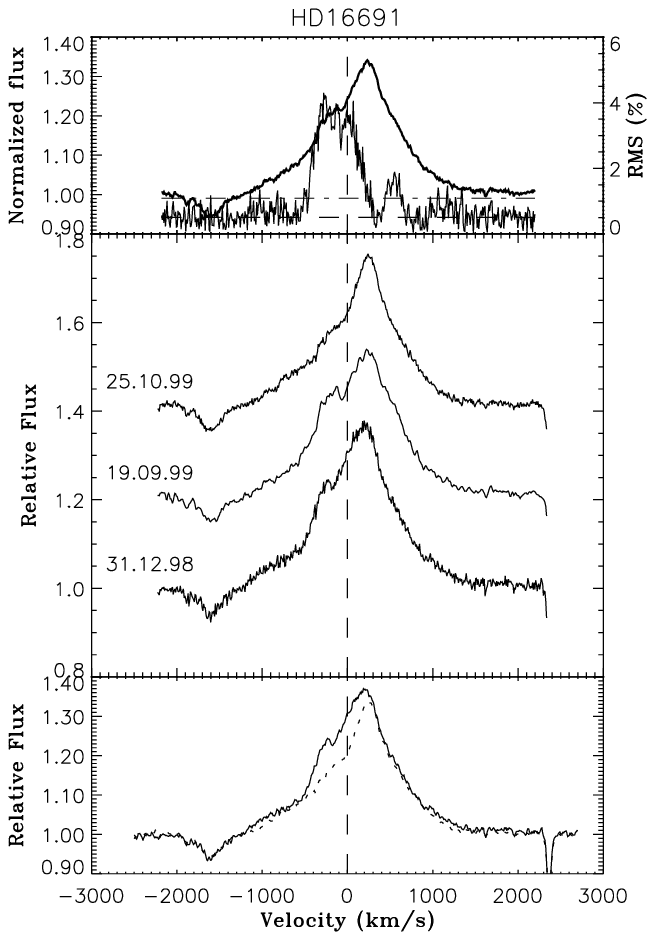}}
\end{minipage}
\hfill
\begin{minipage}{5.9cm}
\resizebox{\hsize}{!}
{\includegraphics{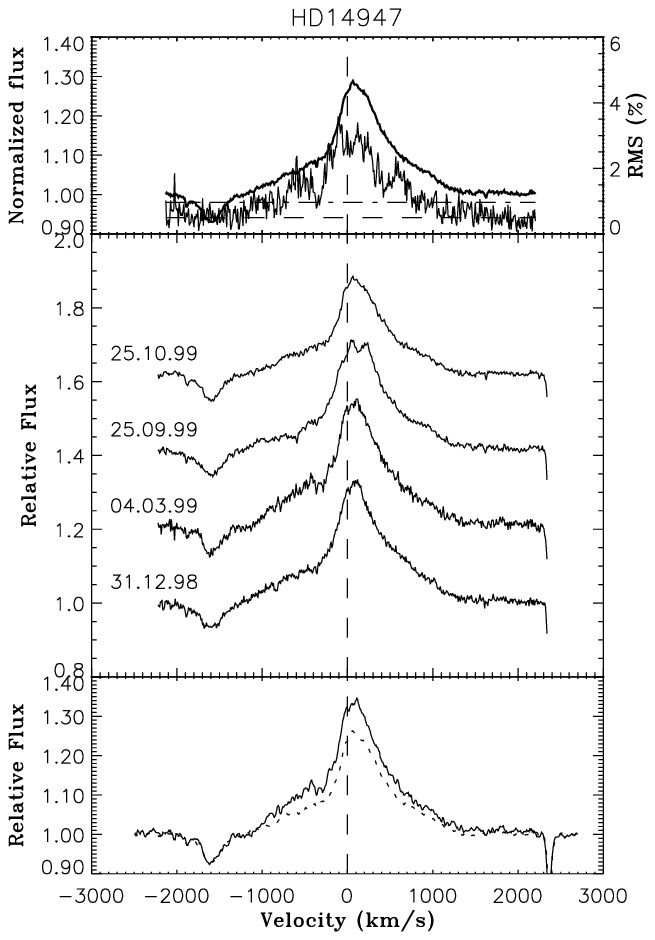}}
\end{minipage}
\caption{Stars of early spectral type (O4/5). {\it Upper part of each
panel:} Mean \Ha profiles (thick line) and $RMS$ deviations as a function of
velocity across the line. {\it Middle part of each panel:} Time-series of
observed \Ha profiles. {\it Lower part of each panel:} \Ha profiles with
maximum and minimum wind emission in the time series.  Velocity scale
centered at the corresponding systemic velocity.}
\label{hots}
\end{figure*}

\subsection{Stars of intermediate spectral type}
\label{ims}

The subset of our supergiant sample with intermediate effective temperatures
($32\,000~{\rm K} < T_{\rm eff} < 37\,000~{\rm K}$) includes six objects
with spectral types ranging from O6 to O8.5. Apart from HD~24\,912,
whose \Ha is in absorption and which will be considered separately at the
end of this subsection, all the other objects exhibit \Ha in emission (see
Fig.~\ref{mids}). 

The observations indicate the presence of noticeable \Ha lpv in any star
of this sub-group (see the middle parts of the panels in
Fig.~\ref{mids}). The emission component varies from sharp, single-peaked
emission to a double-peaked and somewhat broader structure, while the
absorption component changes from a well-defined trough to a weak, hardly
noticeable feature. At the same time, the morphology of the profiles does
not change drastically and always consists of a P~Cygni-like core on top of
extended emission wings.  The emission peak of the mean \Ha profiles is
red-shifted with respect to \vsys, whereas the position of the absorption
dip is different for various stars and ranges from 260 to 360 \kms relative
to the emission peak.  

For the majority of the stars our TVS analysis reveals the presence of {\it
significant} lpv concentrated at the P~Cygni-like core. The variations are
distributed almost symmetrically with respect to the rest wavelength.  Only
in HD~225\,160 and in HD~338\,926 they are concentrated towards the blue and
the red side of the line, respectively. The velocity limits of {\it
significant} variability are always well defined except for HD~210\,839,
where these limits seemed to be more uncertain. 

To clarify the situation, we investigated the individual \Ha profiles of the
HD~210\,839 time-series and found that part of the lpv detected between 
-900 and -500 \kms and between 500 and 700 \kms might be due to imperfect 
telluric line correction. This possibility was taken into account by providing 
two entries for $v_{\rm b}$ and $v_{\rm r}$ of HD~210\,839 in Table~\ref{lpv}.

Interestingly, few years ago \citet{Kaper99} reported evidence for {\it
significant} variability in \Ha of HD~210\,839 concentrated towards the blue
side of the line (between 0 and -400 \kms). The authors attributed this
variability to the cyclic (P = 1.4 days) appearance of DACs in UV resonance
lines. This finding is somewhat different from the one derived by us, where
{\it significant} variability in \Ha of HD~210\,839 extends towards the
emission peak of the line as well. This ``inconsistency'' might be explained
by suggesting that in addition to the short-term variability caused by DACs
there is another variability component that operates on a time scale longer
than covered by the observations used by Kaper et al.

Interestingly as well, the morphology of \Ha of  HD~192\,639 does not indicate
any evidence for deviations from spherical symmetry and homogeneity, in
contrast to the findings by \citet{Rauw01}. If not due to observational
selection, this result might suggest that the wind structure observed by
\citet{Rauw01} was not a permanent but a transient feature of the wind,
similar to the one observed in $\alpha$ Cam \citep{Markova02}.
\begin{figure*}
\begin{minipage}{5.9cm}
\resizebox{\hsize}{!}
      {\includegraphics{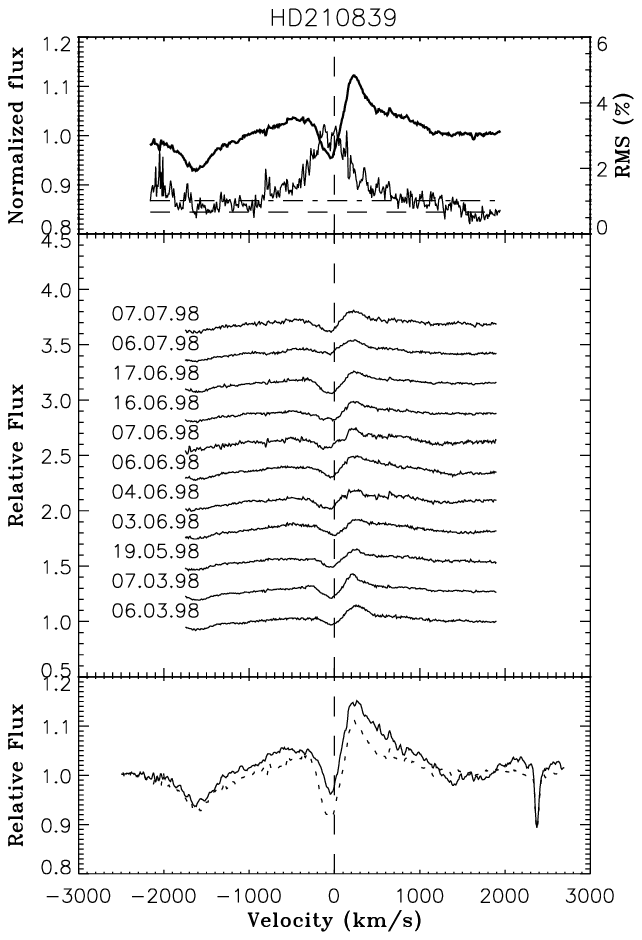}}
\end{minipage}
\hfill
\begin{minipage}{5.9cm}
\resizebox{\hsize}{!}
      {\includegraphics{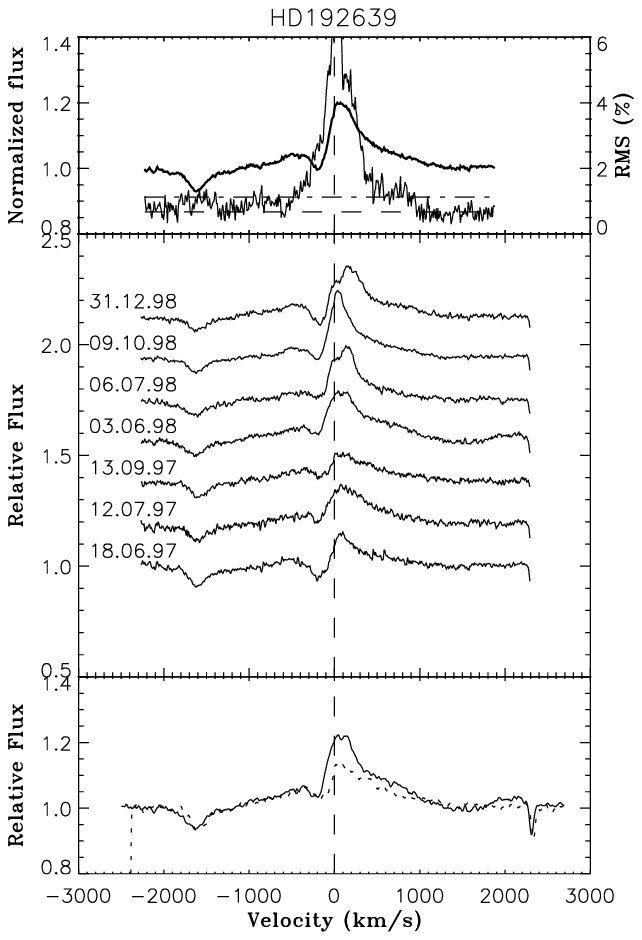}}
\end{minipage}
\hfill
\begin{minipage}{5.9cm}
\resizebox{\hsize}{!}
      {\includegraphics{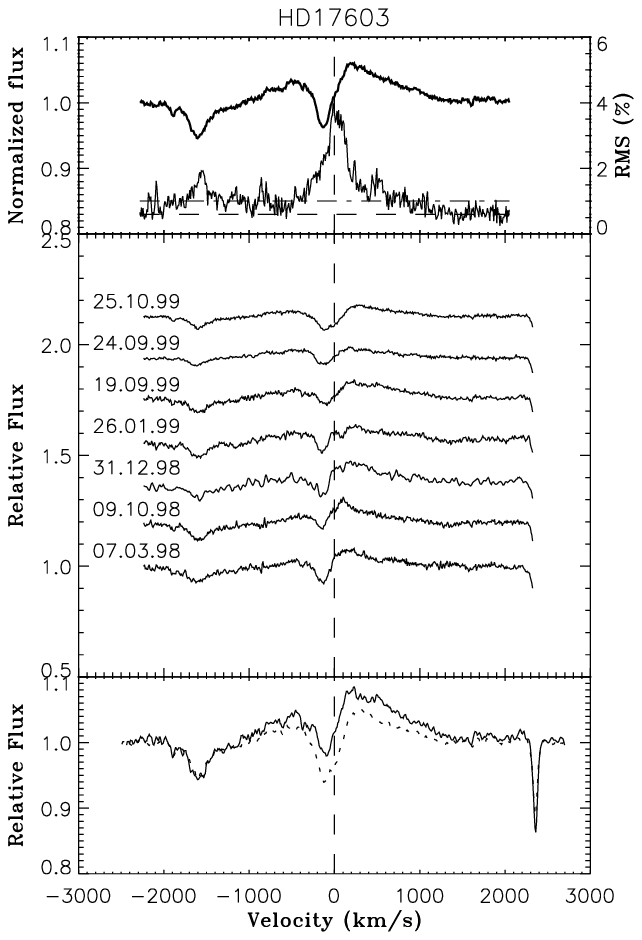}}
\end{minipage}
\hfill

\begin{minipage}{5.9cm}
\resizebox{\hsize}{!}
      {\includegraphics{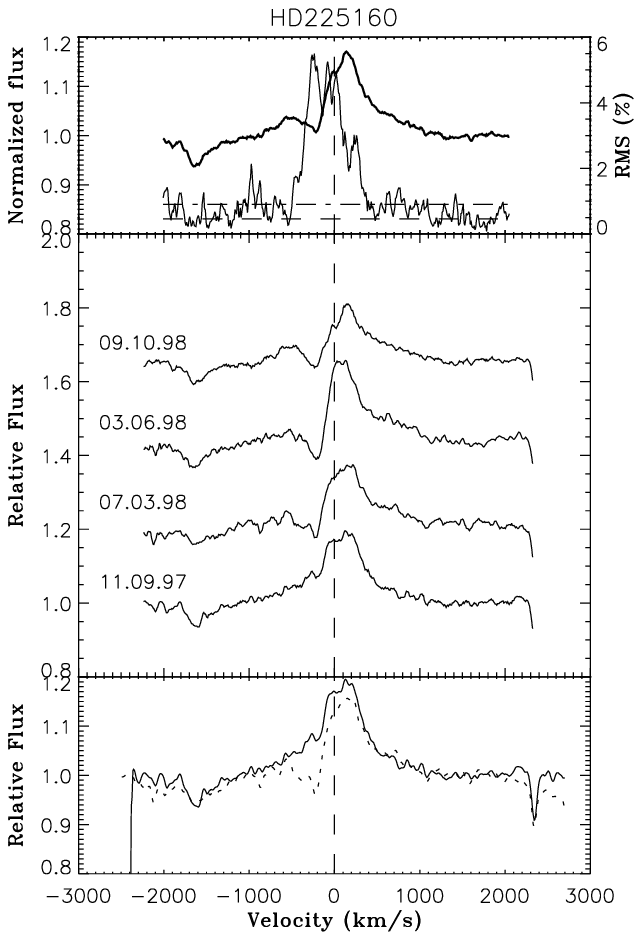}}
\end{minipage}
\hfill
\begin{minipage}{5.9cm}
\resizebox{\hsize}{!}
      {\includegraphics{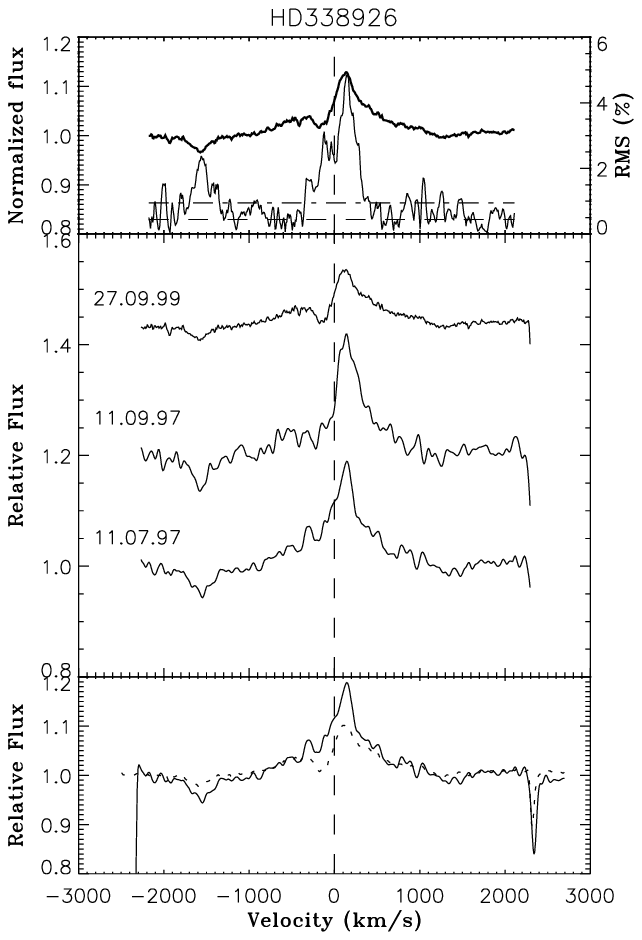}}
\end{minipage}
\begin{minipage}{5.9cm}
\resizebox{\hsize}{!}
      {\includegraphics{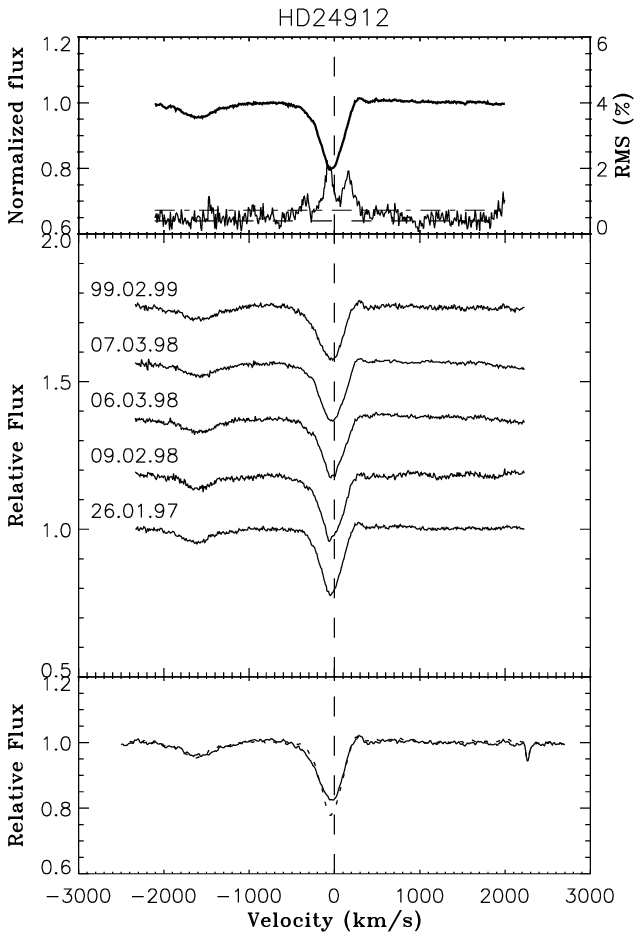}}
      \end{minipage}
\caption{As Fig.~\ref{hots}, but for stars of intermediate spectral type (O6 to O8.5).}
   \label{mids}
\end{figure*}

Except for HD~17\,603 and HD~338\,926 no indication of {\it significant}
absorption lpv was found. Regarding HD~210\,839 and
HD~192\,639 this finding does not agree with results reported by
\citet{Fullerton96} and by \citet{deJong99}.\footnote{\citet{deJong99} have
recognized HD~210\,839 as a non-radial pulsator.} 
Thus, we conclude that at least for four out of the five
stars in this group contributions from photospheric lpv can be expected.

For all stars a genuine variability in the equivalent width of \Ha has been
found. Neglecting the possible contribution from absorption lpv (if present)
we found that the (rather extreme) variations in the \Ha net wind
emission can be accounted for by a 12\% (HD~225\,160) to 44\% (HD~210\,839)
variation in \Mdot. 

\paragraph{\it HD~24\,912 ($\xi$ Per)} was observed once in 1997, three
times in 1998 and once in 1999. The \Ha profiles (lower-right panel of
Fig.~\ref{mids}) appear completely in absorption, in contrast to the
profiles of the other sample stars of similar spectral type. The red wing is
steeper than the blue one and the absorption dip is blue-shifted with 
respect to the stellar rest frame. Profiles with similar signatures are
typical for stars with weaker winds where the \Ha photospheric absorption is
partly filled in by wind emission. 

The TVS of \Ha consists of a blue-shifted, single-peaked component plus a
double-peaked structure with maximum amplitudes concentrated at the
absorption core and at the red extension of the profile. In an absorption
profile, a double-peaked TVS might indicate the presence of radial velocity
variability caused by pulsations \citep{Fullerton96}. Indeed, our
measurements show that the velocity of the absorption core of \Ha varies
around a mean value of -43$\pm$13\kms\footnote{The uncertainty in individual
radial velocity measurements equals to $\pm$4.5~\kms, i.e., half the bin step
of 0.2 \AA.}. Since the position of the absorption dip does not seem to
depend on the line strength (i.e., on the strength of the wind emission) we
suggest that the observed radial velocity variability is mostly (if not
completely) due to changes in the stellar photosphere and probably 
caused by pulsations \citep{deJong99,deJong01}. 

\begin{figure*}
\begin{minipage}{5.9cm}
\resizebox{\hsize}{!}
{\includegraphics{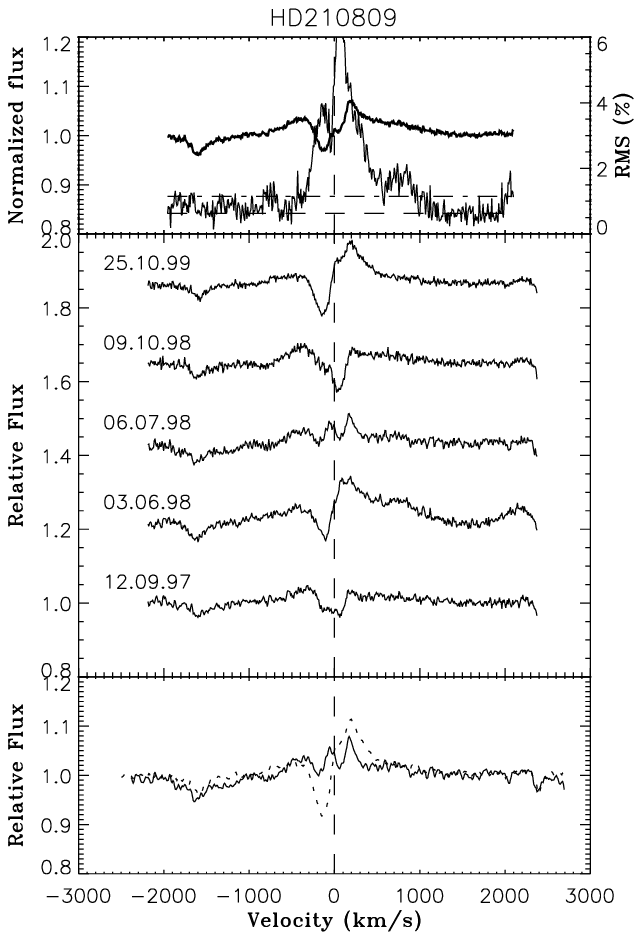}}
\end{minipage}
\hfill
\begin{minipage}{5.9cm}
\resizebox{\hsize}{!}
{\includegraphics{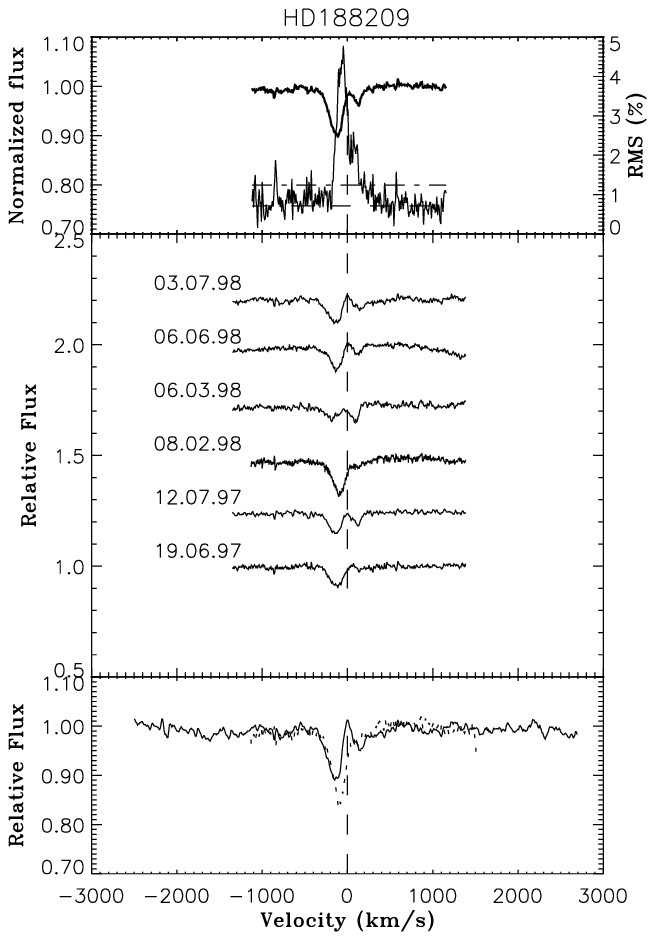}}
\end{minipage} 
\hfill  
\begin{minipage}{5.9cm}
\resizebox{\hsize}{!}
{\includegraphics{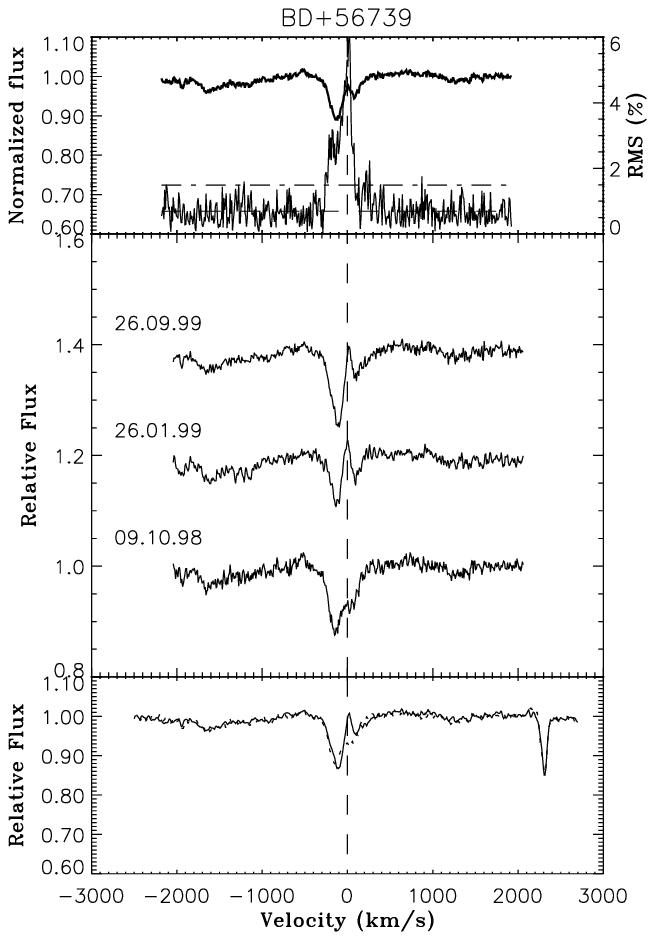}}
\end{minipage}
\hfill

\begin{minipage}{5.9cm}
\resizebox{\hsize}{!}
{\includegraphics{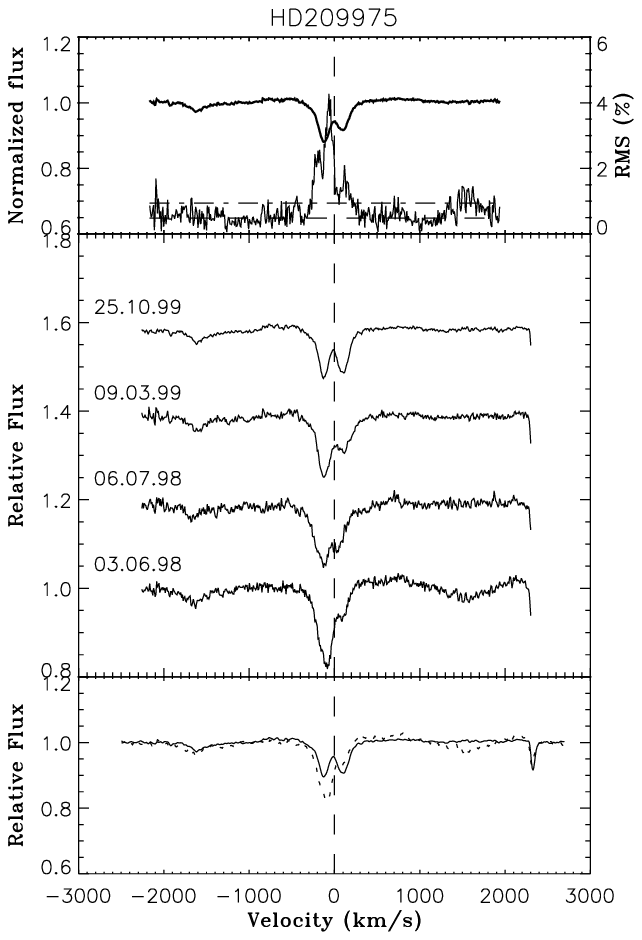}}
\end{minipage}
\hfill
\begin{minipage}{5.9cm}
\resizebox{\hsize}{!}
{\includegraphics{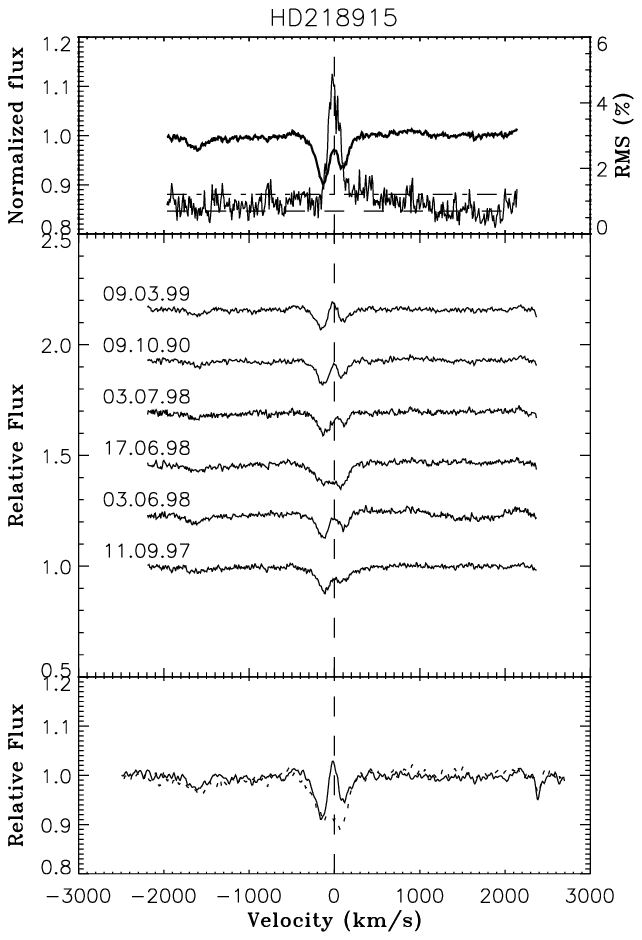}}
\end{minipage}
\hfill
\begin{minipage}{5.9cm}
\resizebox{\hsize}{!}
{\includegraphics{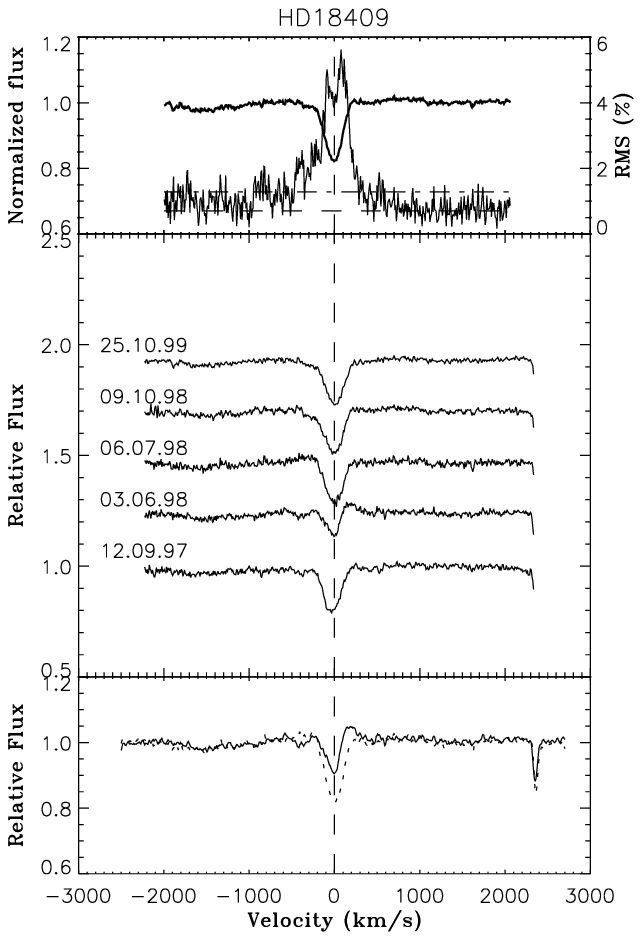}}
\end{minipage}
\caption{As Fig.~\ref{hots}/\ref{mids}, but for stars of
later spectral type (O9 to O9.7). 
Note the single-peaked, blue-shifted absorption profiles observed in
HD~188\,209 in June, 1997 and February, 1998. Profiles with similar shape
have not been observed by \citet{Israelian00} throughout their long-term
monitoring campaign. Note also the P~Cygni-like profile observed in 
HD~18\,409 in June, 1998. This profile is completely different from the rest
of the time series and indicates a strong increase in density in the innermost
wind region.}
\label{lts}
\end{figure*}

In addition to this, another variability component seems to be present in
HD~24\,912, as indicated by the blue-shifted, single-peaked feature in the \Ha
TVS, located between -200 and -450 \kms. This velocity interval is similar
to the interval of the 2-d period variation established by \citet{Kaper97}
and by \citet{deJong01}, and both findings might have the same origin.

The EW of \Ha varies between 1.3 and 1.8 \AA,\, in good agreement with the
limits derived by \citet{Kaper97}. Interpreted as due to variations in the
mean wind density, these limits comply with a $\pm$16\% variation in \Mdot,
which is smaller than the error of the individual \Mdot estimates
($\pm$25\%, \citealt{Markova04}) and thus insignificant.  Again, this
finding is consistent with corresponding results from \citet{Kaper97}. Based
on simultaneous UV and optical observations of $\xi$ Per, the latter authors
suggested that the EW variability of \Ha is caused by the presence of
large-scale, time-dependent structures in the wind.

In summary, we conclude that the variability we have observed in $\xi$ Per
is a mixture of variations originating both from the photosphere (caused by
non-radial pulsations) and from the wind (presumably connected to the
appearance of DACs). 

\subsection{Stars of late spectral type}
\label{lst}

The subset of sample stars of late spectral type (O9 to O9.7, $T_{\rm eff} <
32\,000$~K) includes 6 objects. Five of them show \Ha profiles with 
similar morphology whereas another one - HD~210\,809 - exhibits a 
multitude of differently shaped \Ha profiles and will be considered separately 
at the end of this subsection.

The \Ha profiles of HD~188\,209, BD+56 739, HD~209\,975, HD~218\,915 and 
HD~18\,409 appear in absorption, partly filled in by wind emission (see 
Fig. ~\ref{lts}). The shape of the profiles can vary, from star to star and 
for a given star as a function of time, from double-peaked absorption with a
central reversal peaked at the stellar rest frame to an asymmetric,
blue-shifted absorption feature, with the red wing being steeper than the
blue one. \Ha profiles with similar signatures have been found in early
B-type supergiants as well, e.g., \citet{Ebbets82}.

No indication of extended emission wings has been found in this subgroup,
except for BD~+56\,739.  This object shows weak, but clearly visible
emission wings extending to about $\pm$1200 \kms, suggesting a relatively
strong wind.\footnote{Weak emission wings ($\pm$900 \kms) might be 
present also in HD~209\,975.}

All stars show evidence of real lpv in H$\alpha$. For part of the sample
stars, the deviations are distributed almost symmetrically with respect to
the stellar rest frame, with maximum amplitudes concentrated almost at the
central reversal, whereas for others (e.g. HD~209\,975), the variations are
stronger blueward from the rest wavelength. In the particular case of
HD~209\,975 this blue-to-red asymmetry of the TVS might be explained by
bluewards migrating DACs \citep{Kaper97}. Two entries are given for
HD~18\,409, since the blue velocity limit for {\it significant} \Ha 
variability is somewhat uncertain.

For none of the stars in this group we found evidence of {\it
significant} variability in the \hee absorption line. This finding is
consistent with the results reported by \citet{Fullerton96} for the two
stars in common, HD~188\,209 and HD~209\,975. In contrast,
\citet{Israelian00} have reported evidence of quasi-periodic absorption lpv
(with P = 6.4$^{\rm d}$) for HD~188\,209. Since the data-set used by
\citet{Israelian00} is much more extended than the one used by us and by
\citet{Fullerton96}, we consider their result as more reliable. Thus we
assume that in five out of the six stars the observed variability in \Ha is
dominated by changes in the wind and that only in HD~188\,209 a contribution
from photospheric lpv is to be expected.

Our measurements show that the main source of lpv in \Ha are changes in 
equivalent width. For all stars the \Mdot variations 
needed to account for the extreme changes detected in \We\, exceed the
error of individual determinations and are therefore considered as real. 
As an example, for HD~188\,209 we derived upper and lower limits of 
1.5 and 1.75 x10$^{\rm -6}$\Msun per year, respectively, in full agreement 
with the estimates reported by \citet{Israelian00}. 

\paragraph{\it HD~210\,809.} The individual spectra, shown in the middle
part of the corresponding plot in Fig.~\ref{lts}, indicate the presence of 
dramatic lpv in H$\alpha$. The profiles change from a relatively weak
absorption trough with a flat core via an ordinary/reverse P~Cygni-like
feature to a triple emission structure. In addition, strong emission wings
extending to $\sim$1500~\kms are clearly visible. Line profile
variability with similar signatures has been discussed as an indication 
of long-lived, large-scale wind density perturbation(s), which
co-rotate with the star, giving rise to additional line emission at various
frequencies \citep[ and references therein]{Rauw01}.

As might be expected from the observations, the distribution of lpv in \Ha 
is asymmetric with respect to the rest wavelength, with maximum amplitudes 
concentrated at the P~Cygni-like core. The smaller amplitude deviations
located between 600 and 900 \kms are caused by the appearance of a bump on
the red emission wing in the June 1998 line profile. 

From the TVS of \hee, we found no indication of {\it significant}
photospheric lpv, suggesting that the observed variations in \Ha are caused
by changes in the wind.

Although the observations give clear evidence for deviations from spherical
symmetry, we applied our line-synthesis code (based on a spherical model) to
find constraints on the mass-loss rate variability of the star. Fitting
particularly the first and the last profile of the time series by model
calculations, we found that the extreme variations in \We\, can be
reproduced by variations of $\pm$20\% in \Mdot.

\section{\Ha line-profile variability as a function of stellar and
wind parameters}
\label{Ha_sp}

In order to obtain further clues concerning the origin of wind variability
(as traced by H$\alpha$) in O supergiants, we examined various correlations
between line profile parameters and parameters of the TVS of H$\alpha$, on
the one hand, and fundamental stellar and wind parameters of the sample stars,
on the other. To search for such correlations, we used the Spearman
rank-order correlation test, described, for example, by \citet{Press92}. The
main advantage of this test is that in addition to the correlation
coefficient (more precisely, the linear correlation coefficient of ranks) it
also calculates the statistical significance of this correlation (expressed
as the two-sided significance of its deviation from zero), without any
assumption concerning the distribution of uncertainties in the individual
quantities.  

\subsection{\Ha profile shape as a function of spectral type}
\label{profileshape}

An inspection of the {\it mean} \Ha profiles of the sample stars (all
supergiants!), displayed in Figures~ \ref{hots} to \ref{lts}, shows
that these profiles evolve as a function of spectral type from a slightly
asymmetric emission with a peak value red-shifted with respect to \vsys,
via an emission feature with a P~Cygni-like core, to a feature in
absorption (with or without central emission reversal). In stars of early and
intermediate spectral type extended emission wings can be seen, while in
stars of late spectral type the presence of such wings is rare. 

There are two stars that deviate from this behaviour, HD~24\,912 and 
HD~210\,809. The former one exhibits a pure absorption profile instead of a
P~Cygni-like profile (see Fig.~\ref{mids}). Consequently, its \Ha line 
resembles much more those profiles from luminosity class III than from
luminosity class I objects of the same spectral type. As we have already
pointed out in Paper~I, the parameters of HD~24\,912 are somewhat insecure
due to the uncertain distance - HD~24\,912 is not a member of PerOB2 but a
runaway star \citep{Gies87}. Thus it is rather likely that the discrepancy
in profile shape is due to an erroneous (re-)assignment in luminosity class
\citep{Herrero92} and suggest that the original value as assigned by
\citet{Walborn73}, luminosity class III, is more appropriate (see also
\citealt{Repo04}).

The second outlier, HD~210\,809, shows a P~Cygni-like profile instead of an 
absorption profile partly filled in by wind emission. This is the only star
in the sample for which our observations suggest a strong deviation from
spherical symmetry. From the similarity to the \Ha and He{\sc ii}~4686 
time-series of HD~192\,639 observed and discussed by \citet{Rauw01} (single
and double peak structure in emission), we speculate that also here a
``confined co-rotating wind'' is present. This interpretation, if correct,
would explain the ``peculiar'' shape of the mean \Ha profile derived by us.
Hereafter, we will refer to HD~24\,912 and HD~210\,809 as to ``peculiar''
stars. 

The observed evolution of \Ha in O-supergiants with spectral type (actually
with \Teff) is in fair agreement with results from theoretical line-profile
computations performed in terms of NLTE, spherically symmetric, smooth
stellar wind models, although the strength of the observed P~Cygni-like core
cannot be reproduced in most cases (e.g., \citealt{Repo04}).\footnote{This
problem has not been solved satisfactorily so far.  In particular, it is
still unclear if the blue-shifted absorption core is solely related to the
He{\sc ii} blend or other effects (clumps, deviations from 1-D geometry)
have to be accounted for additionally.} The main drivers of this evolution
are: decreasing line emission caused by decreasing wind density (since \Mdot
decreases with decreasing \Teff and \logl, see \citealt{Vink00}) and
decreasing contribution of the \he blend.  (In contrast, purely photospheric
\Ha profiles of a given luminosity class do not change significantly as a
function of \Teff , because in this temperature regime the photospheric 
ionization fraction of neutral hydrogen remains fairly constant.) Outliers
can occur either as a result of strong deviations from spherical symmetry
and homogeneity in the wind (due to, e.g., fast rotation, CIRs, clumps) or
as a result of an erroneous spectral type/luminosity class classification or
uncertain/wrong parameters.

\begin{figure*}
\begin{minipage}{8.5cm}
\resizebox{\hsize}{!}
{\includegraphics{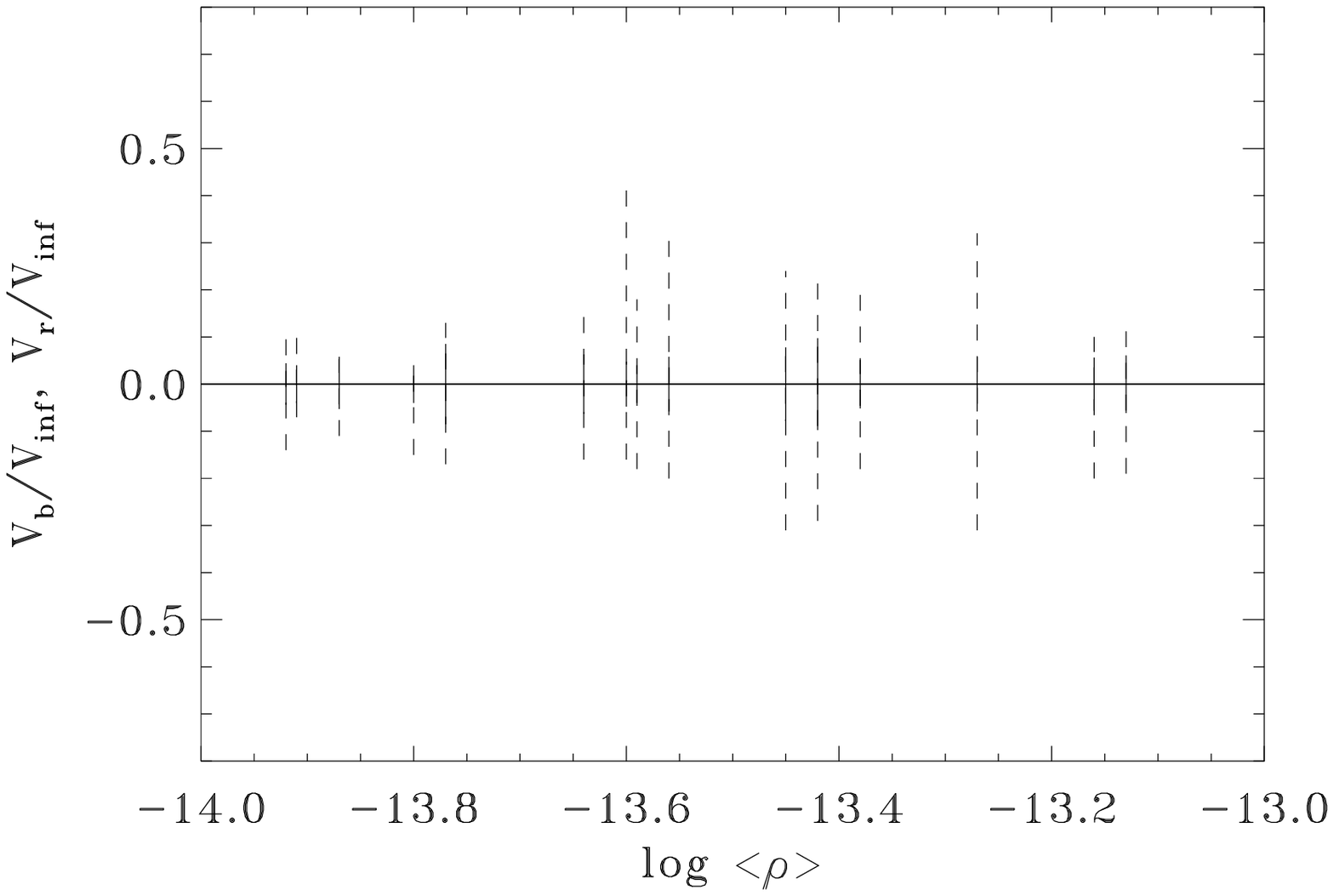}}
\end{minipage}
\hfill
\begin{minipage}{8.5cm}
\resizebox{\hsize}{!}
{\includegraphics{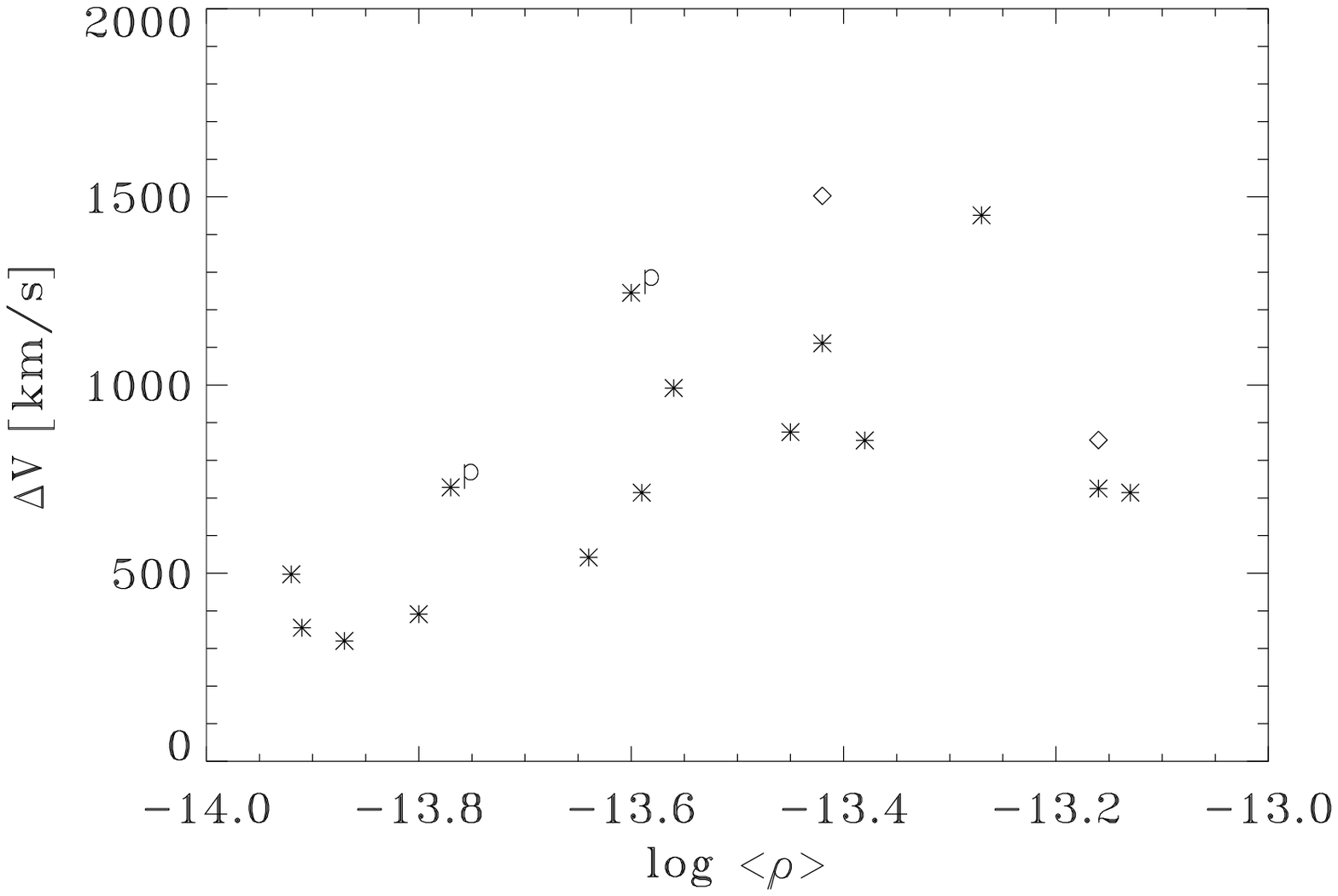}}
\end{minipage}
\hfill
\caption{Blue and red velocity limits (left panel) and velocity width (right
panel) of {\it significant} lpv in H$\alpha$, as a function of mean wind density 
of the sample stars. Thick vertical lines denote the corresponding projected
rotational velocity, $\pm v \sin i$.  Asterisks refer to the conservative
estimates for $\Delta V$, while diamonds mark the non-conservative
ones.}
\label{lpv_1}
\end{figure*}

\subsection{Red-shifted emission-peaks} 

Our observations suggest that the position of the emission peak of the \Ha
profiles of O supergiants depends on the strength of the wind: for stars with 
weaker winds (\Ha in absorption with/without central reversal) this peak is
centered almost at the rest wavelength, whereas for stars with stronger winds
(\Ha in emission) it is red-shifted instead. This observation is supported
by results of the correlation analysis, which shows that the velovity of the 
emission peak,\ve\, correlates significantly with $<\rho>$, (.79/0.0007) and 
in addition with \Teff (.82/0.0003).\footnote{In this particular case and 
because of the reasons outlined above, the ``peculiar'' stars HD~210\,809 
and HD~24\,912 have been discarded from the correlation analysis.}

Hereafter, numbers in brackets denote the Spearman rank correlation
coefficient and the two-sided significance of its deviations from zero.
Since the latter quantity measures the probability to derive a given
correlation coefficient from uncorrelated data, smaller values mean higher
significance of the correlation.
 
Red-shifted emission peaks have been observed in UV resonance lines of 
O-type stars, where this finding (in parallel with the presence of an
extended absorption trough) can been explained in terms of
``micro-turbulence'' effects, with $v_{\rm micro}$ of the order of 0.1 \vinf
(e.g., \citealt{Hamann80} and references therein, \citealt{Groene89}). On a
first glance, the phenomenon seen in \Ha seems to be somewhat similar. In
contrast to the situation for UV resonance lines, however, our \Ha profile
simulations meet no problem in reproducing the red-shifted peak, even if the
shift is large, {\it without any inclusion of micro-turbulence}. This can be
seen clearly by comparing theoretical profiles with observations, e.g.,
\citet{Markova04,Repo04}. A closer inspection of the profile formation
process reveals that the apparent shift of the emission peak results (at
least in our simulations) from the interaction between the red-side of the
Stark-broadened photospheric profile and the wind emission. Let us note that
we do not exclude the presence of micro-turbulence but that we simply do
not need it to reproduce the observed amount of \ve.  

\begin{table*}
\caption {\Ha line profile and variability parameters. $N$ denotes the
number of available spectra.  \ve\, is the velocity of the emission peak
while $v_{\rm b}$, $v_{\rm r}$ are the ``blue'' and ``red'' velocity limits
of significant variability. All velocity data are measured with respect to
the stellar rest frame and given in \kms. $\sigma_0$ is the standardized
dispersion of the corresponding time-series. $r_{\rm max}$ (expressed in \Rstar)
denotes the upper limit in physical space where significant variations in
\Ha are present. \We\, is the mean equivalent width of net wind emission and
its standard deviation, both given in \AA. $W_{\rm phot}$ is the equivalent
width of the photospheric component. \Mdot$_{\rm min}$ and \Mdot$_{\rm max}$
(in $10^{-6}$ \Msun/yr) denote the corresponding limits if the observed variability 
is attributed to
variations in \Mdot alone, while $\Delta$\Mdot is the amplitude of this
variability expressed in percents of \Mdot$_{\rm min}$. $<\rho>$ is the mean
wind density (Eq.~\ref{rhobar}), and $A_{\rm N}$ and $a_{\rm N}$ are 
the mean and the fractional amplitudes (Eqs.~\ref{A_N} and \ref{a_lpvN}), respectively.} 
\label{lpv}
\tabcolsep1.0mm
\begin{tabular}{lrrclcccrrlrccr}
\hline
\hline
\multicolumn{1}{c}{Object}
&\multicolumn{1}{c}{\#}
&\multicolumn{1}{c}{$N$}
&\multicolumn{1}{c}{lpv(a)}
&\multicolumn{1}{c}{\ve}
&\multicolumn{1}{c}{$\sigma_0$*100.}
&\multicolumn{1}{c}{[ $v_{\rm b}, v_{\rm r}$]}
&\multicolumn{1}{c}{$r_{\rm max}$}
&\multicolumn{1}{c}{\We/$W_{\rm phot}$}
&\multicolumn{1}{c}{\Mdot$_{\rm min}$}
&\multicolumn{1}{c}{\Mdot$_{\rm max}$}
&\multicolumn{1}{c}{$\Delta$\Mdot} 
&\multicolumn{1}{c}{$log <\rho>$}
&\multicolumn{1}{c}{$A_{\rm N}$}
&\multicolumn{1}{c}{$a_{\rm N}$}
\\
\hline	

HD 190\,429 &1 &  9 &  no &  210 &0.93 & [ -491, 234] & 1.23 & 10.99$\pm$0.63/3.23 &
13.0 &  14.0 &  8\% &   13.16 & 2.24 &   6.07$\pm$  0.03\\

& &  &  &  & & [-491, 363]  & & &
&  &  &  & 2.09 &  5.73$\pm$0.03\\

HD 16\,691 &2 &  3 &  no &  223 &0.51 & [ -448, 266] & 1.22 & 10.87$\pm$0.81/3.23 &
12.0 &  13.0 &  8\% &   13.13 & 2.89  
&   7.17$\pm$  0.02\\

HD 14\,947 &3&  4 &  no &   82 &0.50 & [ -704, 747] & 1.43 &  9.35$\pm$0.90/3.24 &
14.5 &  16.0 & 10\% &   13.27 & 1.90 
&   7.84$\pm$  0.02\\

HD 210\,839 &4& 11 & yes &  220 &0.67 & [ -633, 478] & 1.40 &  4.74$\pm$0.55/3.00 &
6.8 &   9.8 & 44\% &   13.42 & 1.92 
&  11.63$\pm$  0.03\\

& &  &  &  & & [-838, 665] & 1.61  & &
&  &  &  & 1.73 
& 12.95$\pm$0.04  \\

HD 192\,639 &5&  7 &  no &  140 &0.67 & [ -393, 460] & 1.27 &  6.15$\pm$0.52/3.02 &
4.7 &   5.4 & 16\% &   13.38 & 3.36  
&  13.77$\pm$  0.03\\

HD 17\,603 &6&  7 & yes &  210 &0.60 & [ -373, 619] & 1.27 &  4.04$\pm$0.49/2.84 &
5.5 &   7.2 & 31\% &   13.56 & 1.87 
&  12.49$\pm$  0.05\\

HD 24\,912 &7&  5 & yes &  -50 &0.40 & [ -419, 309] & 1.12 &  1.36$\pm$0.14/2.85 &
4.5 &   5.2 & 16\% &   13.77 & 1.09 
&  15.32$\pm$  0.14\\

HD 225\,160 &8&  4 &  no &  140 &0.46 & [ -491, 384] & 1.33 &  4.51$\pm$1.04/2.62 &
5.1 &   5.7 & 12\% &   13.45 & 3.21 
&  15.61$\pm$  0.02\\

HD 338\,926 &9&  3 & yes &  140 &0.44 & [ -352, 362] & 1.21 &  4.54$\pm$0.38/2.61 &
4.5 &   5.4 & 20\% &   13.59 & 2.48 
&  11.78$\pm$  0.03\\

HD 210\,809 &10&  5 &  no &   10 &0.63 & [ -336, 909] & 1.15 &  3.63$\pm$0.63/2.59 &
3.2 &   4.5 & 41\% &   13.60 & 2.77 
&  24.11$\pm$  0.05\\

HD 188\,209 &11&  6 & yes &   -6 &0.72 & [ -176, 144] & 1.09 &  1.56$\pm$0.19/2.18 &
1.5 &   1.8 & 17\% &   13.87 & 2.67 
&  16.37$\pm$  0.08\\

BD+56\,739 &12&  3 &  no &   13 &0.69 & [ -295,  96] & 1.12 &  1.78$\pm$0.31/2.18 &
2.1 &   2.5 & 19\% &   13.80 & 2.84 
&  24.66$\pm$  0.72\\

HD 209\,975 &13&  4 &  no &   26 & 0.48 & [ -277, 220] & 1.09 &  1.36$\pm$0.29/2.18 &
1.5 &   1.9 & 27\% &   13.92 & 1.90 
&  20.85$\pm$  0.10\\

HD 218\,915 &14&  6 &  no &   -8 &0.70 & [ -149, 206] & 1.07 &  1.81$\pm$0.24/2.18 &
1.6 &   2.0 & 25\% &   13.91 & 2.51 
&  17.14$\pm$  0.10\\

HD 18\,409 &15&  5 &  no &    3 &0.70 & [ -274, 268] & 1.08 &  1.40$\pm$0.47/2.17 &
1.5 &   2.2 & 47\% &   13.64 & 3.39 
&  47.18$\pm$  0.11\\
\hline
\end{tabular}
\end{table*}

\begin{figure*}
\begin{minipage}{8.5cm}
\resizebox{\hsize}{!}
{\includegraphics{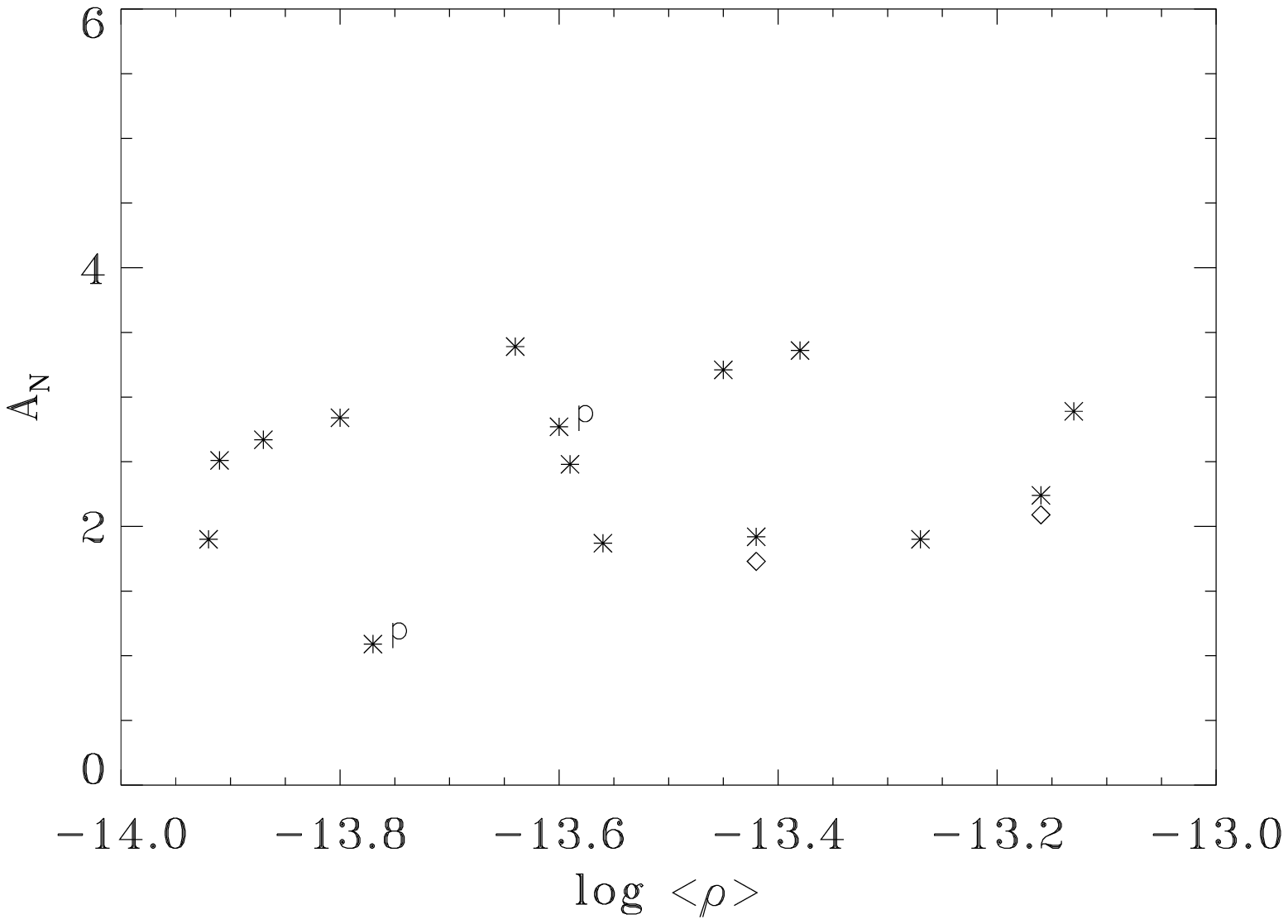}}
\end{minipage}
\hfill
\begin{minipage}{8.5cm}
\resizebox{\hsize}{!}
{\includegraphics{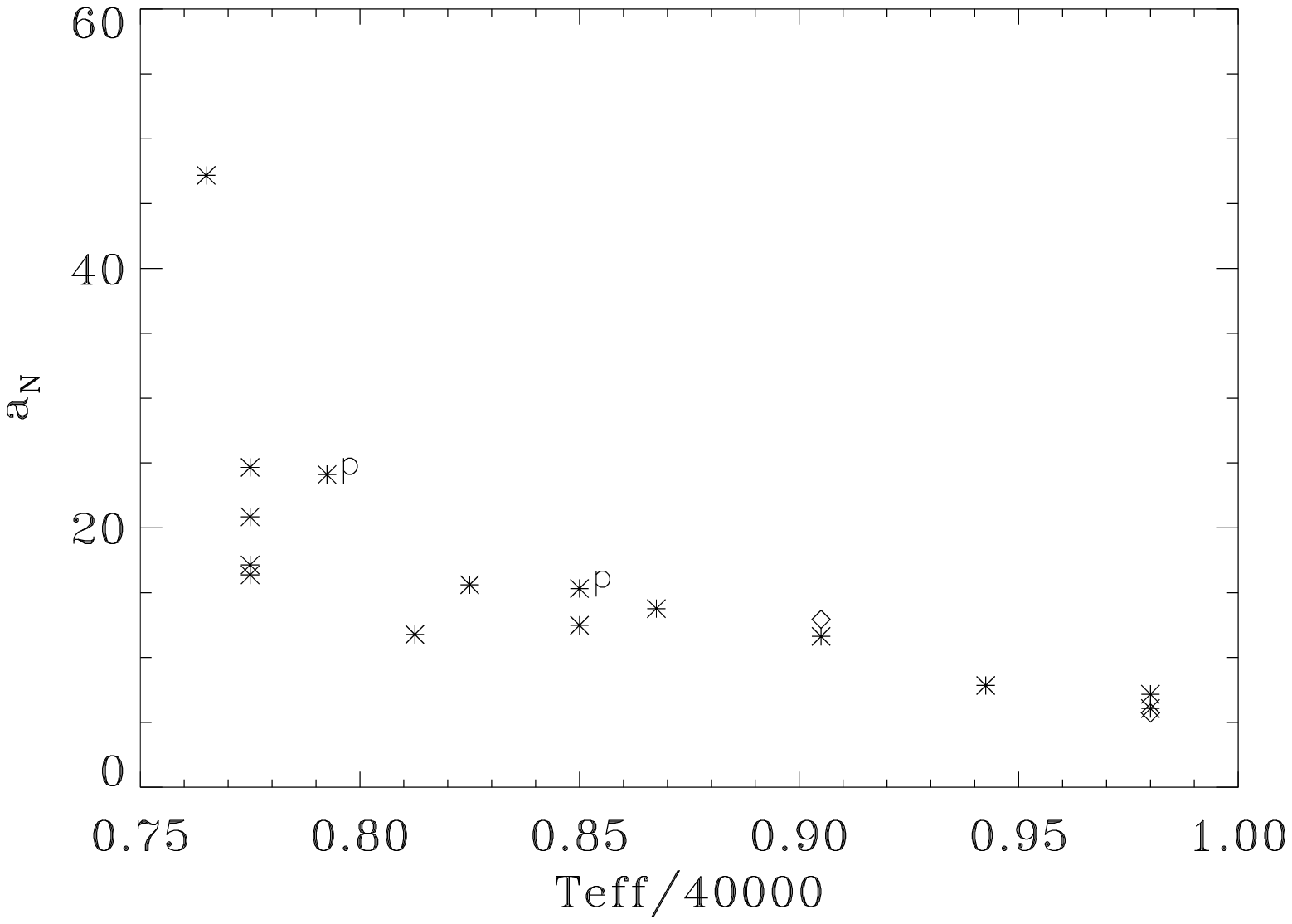}}
\end{minipage}
\hfill
\begin{minipage}{8.5cm}
\resizebox{\hsize}{!}
{\includegraphics{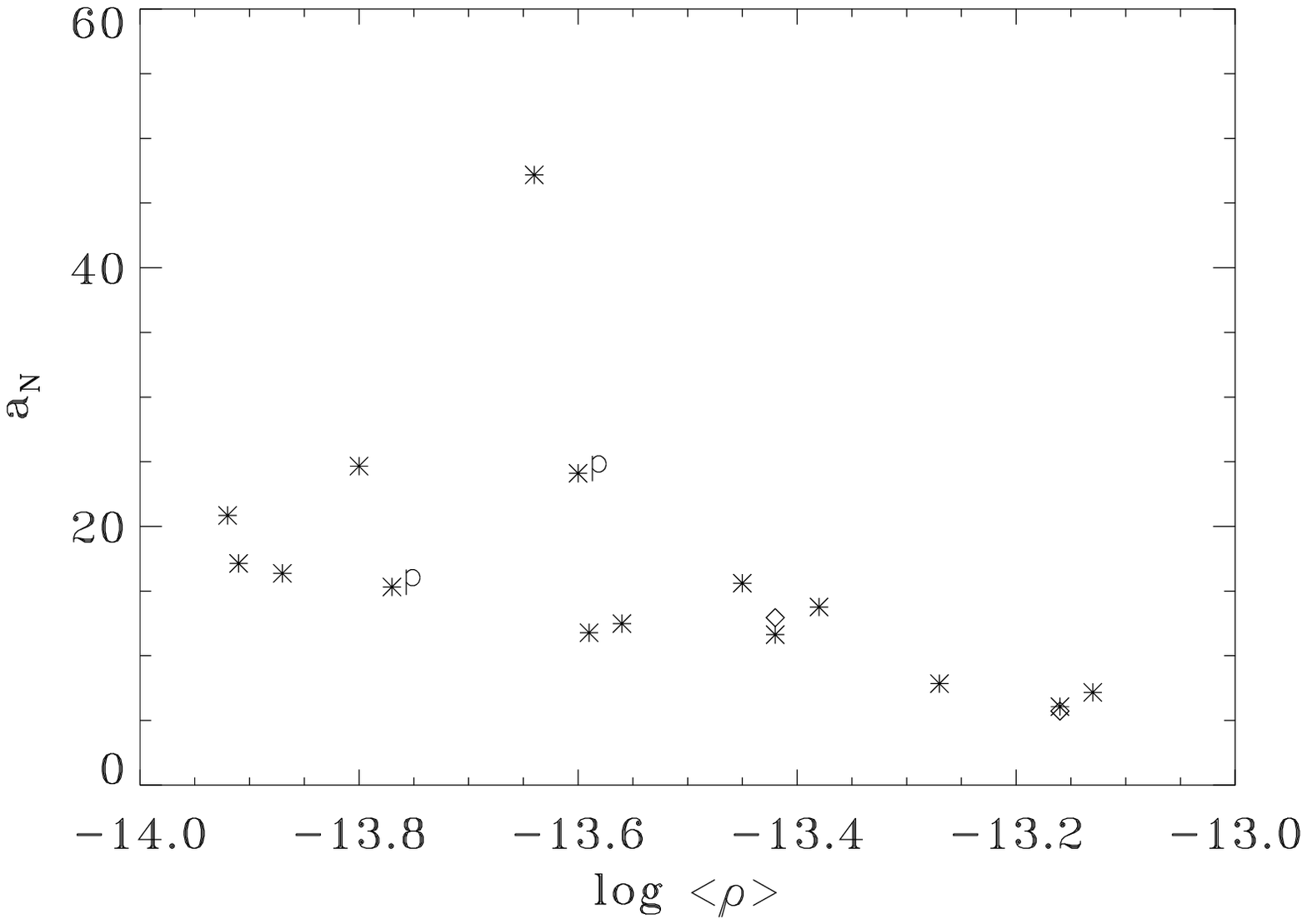}}
\end{minipage}
\caption{Examples of significant correlations between the fractional 
amplitude of deviations, $a_{\rm N}$, as defined in the present study, and
stellar and wind parameters of the sample stars. Distributions of the mean
amplitude $A_{\rm N}$ are shown for comparison.
Asterisks refer to the conservative estimates, while diamonds mark the
non-conservative ones. Positions of the two ``peculiar'' stars are denoted
by `P'.}
\label{lpv_2}
\end{figure*}

\subsection{Properties of the TVS as a function of stellar and wind
parameters}
\label{tvs} 

Before investigating the properties of the \Ha TVS for our sample stars,
let us point out that all results outlined below refer to the ``conservative
case'' (see Sect.~\ref{method}).  Although the ``non-conservative'' data have
not been analyzed in detail, they are included in the corresponding plots
and we will comment on their influence on the final outcome.

\subsubsection{Distribution of \Ha line-profile variability in velocity space}
\label{vel_ext}

In Figure~\ref{lpv_1} we show the blue and red velocity limits (left panel,
dashed) and the velocity width $\Delta V$ (absolute value, right panel) of
\Ha lpv as a function of the mean wind density. Thick vertical lines
correspond to the projected rotational speed, $\pm v \sin i$. Asterisks
refer to the conservative estimates for $\Delta V$, while diamonds mark 
the non-conservative cases. In combination with the results described in
Sect.~\ref{Ha_time}, these figures indicate that: 
\ben
\item[i)] for all stars the \Ha lpv extends beyond the limits determined by 
stellar rotation and thus must be linked to the wind;

\item[ii)] in most of our sample stars the variations occur either 
symmetrically (within the error) with respect to the rest wavelength or with
a weak blue-to-red asymmetry. For two objects, HD~17\,603 and
HD~210\,809, the TVS has a noticeable red-to-blue asymmetry, 
while in other two stars, HD~190\,429A and HD~16\,691, the
variations are stronger and more extended bluewards of the rest wavelength.

\item[iii)] the velocity width for {\it significant} variability in \Ha 
is larger in stronger winds than in weaker ones. 
There are two stars that deviate from this rule: HD~190\,429A and 
HD~16\,691 which exhibit variations over a velocity interval that is 
considerably smaller than expected from the strength of their winds. 
\een

Further analysis of the velocity data listed in Table~\ref{lpv} shows that
in all sample stars {\it significant} lpv in \Ha occurs below 0.3 \vinf.
Converted to physical space, this yields an upper limit of roughly 1.5
\Rstar for the observed variability. Additionally, we found a significant
correlation between $r_{\rm max}$ and $\log <\rho>$, (0.87/0.00001). This
result as well as the possible dependence between $\Delta V$ and $\log
<\rho>$ (0.62$\pm$0.01) are readily understood in terms of an increasing
wind volume which contributes to the \Ha emission, as a function of wind
density.

We are aware of the fact that due to observational selection effects and
other uncertainties affecting the determination of the velocity limits, the
results described above might be questioned. However, note that: (i) the
probability to obtain a symmetric TVS for a star with a strongly asymmetric
wind, using snapshot observations, is very low; (ii) the established 
correlations between $\Delta V$ and $\log <\rho>$ (right panel of
Figure~\ref{lpv_1}) and between $r_{\rm max}$ and $\log <\rho>$ are both
physically reasonable, which in turn supports the reliability of the limits
determined by us; (iii) although the limits and hence $\Delta V$ changes when
non-conservative instead of conservative estimates are considered, the final
outcome does not change (see Figure~\ref{lpv_1} and Table~\ref{lpv}). 

Thus, we assume that the velocity limits determined here do not seem to be
strongly biased by either observational selection or uncertainties in the
measured quantities. (But see also the next sub-section.)

If a good temporal resolution of the variability time-scale is provided, the
velocity distribution of lpv in \Ha will allow to obtain significant
information about the wind geometry \citep{Harries}. On the other hand, and
as we shall show later on (see Sect~\ref{simul}), even in the case of
snapshot observations some hints about wind structures can be derived.

\subsubsection{Mean and fractional amplitudes of deviations as a function 
of stellar and wind parameters} 
\label{mean_amp}

Our TVS analysis shows that the mean amplitude of deviations always exceeds 
the corresponding threshold for {\it significant} variability, indicating 
genuine variability in H$\alpha$. The actual values of $A_{N}$ range between
1 and 4 percent of the continuum flux, without any clear evidence for 
dependence on stellar and wind parameters of the sample stars. This result 
is illustrated in the upper left panel of Fig.~\ref{lpv_2}, where the
estimates for $A_{N}$ are shown vs. $\log <\rho>$. 

This independence of $A_{N}$ on $\log <\rho>$ can have a twofold
interpretation: First, {\it if} the mean amplitude is a reasonable measure
for wind variability, then the wind variability is actually more or less
independent on wind-strength. Second, the mean amplitude is not the
appropriate tool to compare the strengths of \Ha lpv. 

Leaving aside these two possibilities, let us first consider the following
problem. If we assume that the sources of observable variability are
distributed over a certain volume which increases as a function of mean wind
density (as it is suggested from the increase of $r_{\rm max}$ with $\log
<\rho>$), then one should expect that also $A_{N}$ should increase with mean
density, since the numerator of this quantity (the integral over
$(TVS-\sigma_0^2)^{0.5}$) increases as a function of the emitting {\it
volume}, whereas the denominator corresponds to an (increasing) 1-D quantity
only.  The fact that the {\it observed} mean amplitude is actually
independent on $\log <\rho>$ shows that such a simple model is not
sufficient to explain the observations. We will come back to this point
again in Sect.~\ref{simul}.

In contrast to the established independence of the mean amplitude on wind
density, our analysis shows the presence of a {\it negative} correlation
between the {\it fractional} amplitude of deviations, $a_{\rm N}$, and a
number of stellar/wind parameters. Scatter plots for the strongest 
correlations, with \Teff (0.92/0.000001) and with $\log <\rho>$
(0.80/0.0003), are illustrated in Fig.~\ref{lpv_2}. In particular, the
decrease of $a_{\rm N}$ with increasing $\log <\rho>$ suggests that the
observed variability per unit fractional net emission is smaller in denser
winds than in thinner ones.

The reliability of the derived values of the mean and fractional
amplitudes has been checked in two ways: {\bf first}, we examined the
stability of the results against effects caused by observational
selection and other uncertainties in the measured quantities;
{\bf second}, we checked the validity of the assumptions underlying our
definitions of $A_{\rm N}$ and $a_{\rm N}$. 

In particular, to clarify to what extent these quantities (and again $\Delta
V$) might be influenced by observational selection effects, we proceeded as
follows: 
\ben
\item[i)]
For the two stars with longest time series we reduced the number of spectra 
by about 30\% while keeping the two most ``extreme'' profiles. The effect of
this data manipulation on the TVS parameters was found to be insignificant. 
\item[ii)]

For the stars with longer time series ($N > 6$) we removed: (a) the
spectrum with minimum \Ha emission; (b) the spectrum with maximum \Ha
emission and (c) the two spectra with minimum and maximum wind emission 
and re-calculated the TVS. In all the three cases the new estimates of
$\sigma_0$ and $\Delta V$ turned out to be quite similar to the original
ones: the established differences were less than a few percent. At the same
time the TVS amplitudes changed by less than 10 percent for stars with \Ha 
in emission and by about 15 to 20 percent for stars with \Ha in (partly
re-filled) absorption. The effect of these changes was again insignificant
concerning the final results for $\Delta V$, $A_{\rm N}$ and $a_{\rm N}$.
\een

Let us finally point out that the estimates of the TVS parameters are
expected to depend strongly on the quality of the data used (i.e., on
$\sigma_0$). To simulate higher S/N (about two times higher than the
original values) we smoothed the spectra in each time series, using a boxcar
average with a width of 4 pixels, and analyzed them in the same way as the
original spectral series. Interestingly, while the effect of this data
manipulation on the estimates of $\Delta V$ was surprisingly small (less
than $\pm$12 percent of the original values), the reaction of the TVS
amplitudes turned out to be quite strong: both $A_{\rm N}$ and $a_{\rm N}$, 
averaged over the whole sample, decreased by 37 and 33 percent,
respectively, compared to their original values.\footnote{Part of this
decrease might be due to the fact that the contribution of higher frequency
variability (if any) has been reduced by smoothing.} Most importantly,
however, the new estimates of $\Delta V$, $A_{\rm N}$ and $a_{\rm N}$ were
found to obey similar dependences on $\log <\rho>$ as implied by the
original data set.\footnote{This result might no longer be true if large
differences in the quality, i.e., in the standardized dispersion of the
spectral time-series, $\sigma_0$, were present (the requirement of
homogeneity).}

Summarizing we conclude that the derived independence of $A_{\rm N}$ on 
stellar and wind parameters as well as the negative correlations between
these parameters and $a_{\rm N}$ cannot be explained in terms of either
observational selection or uncertainties in the measured quantities.

\section{Simulations of lpv in \Ha}
\label{simul}

In order to account for the systematic difference in the strength of \Ha as
a function of spectral type/mean wind density, in Sect.~\ref{method} we
optimized the fractional amplitude of deviations by normalizing the integral
over the TVS to a quantity which we called fractional emission equivalent
width, FEEW (see Eq.~\ref{a_lpvN}). 

The parameter $a_{\rm N}$ defined in this way is thus a measure for the
observed variability (represented by the corresponding TVS) per unit
fractional wind emission and has been used in the previous section to
investigate the dependence of the observed variability on wind density. The
results obtained might be interpreted as an indication that denser winds are
less active than thinner winds, a finding which would give firm
constraints on present hydrodynamical simulations.

Note, however, that (i) our definition of $a_{\rm N}$ implicitly assumes 
that the TVS amplitude is proportional to the corresponding amount of
wind-emission and that (ii) this assumption has not been checked so far. In
particular, if this assumption was justified, the derived values of $a_{\rm
N}$ would provide a robust measure for the ``observed'' degree of
wind-variability, as it is true for the {\it photospheric} lpv's described in
terms of $a_{\rm F}$. 

\subsection{1-D model simulations}
\label{1D_simul}

\begin{table}
\caption {Summary of simple 1-D simulations. Models denoted by``SS'' refer
to spherical shells, models denoted by ``BS'' to broken shells,
respectively.} 
\label{1dmodels}
\begin{tabular}{lll}
\hline
\hline
\\
Series &  properties  & max$(\delta \rho / \rho_0)$ \\
\hline	
SS1 & $\delta v = 0.5 v_{\rm th}$(H)  & $\pm$0.7 \\
SS2 & $\delta v = 1.0 v_{\rm th}$(H)  & $\pm$0.7 \\
BS11 & $\delta m$ = const, $\Delta p$(core) = 0.1 \Rstar & $\pm$0.35 \\
BS12 & $\delta m$ = const, $\Delta p$(core) = 0.1 \Rstar & $\pm$0.7 \\
BS21 & $\delta m$ = const, $\Delta p$(core) = \Rstar & $\pm$0.35 \\
BS22 & $\delta m$ = const, $\Delta p$(core) = \Rstar & $\pm$0.7 \\
SS3 & $\delta m$ = const    & $\pm$0.35 \\
\hline
\end{tabular}
\end{table}

Thus, a test of our hypothesis is urgently required. Ideally, such a
test would make use of at least 2-D models of instable winds, since the
assumption of 1-D shells most probably overestimates the actual
degree of variability. 

Since 2-D simulations involving a consistent physical description are just
at their beginning \citep{Dessart03} and since, to our knowledge, even for
somewhat simpler multi-dimensional models no investigation concerning the
dependence on wind parameters is available (Sect.~\ref{3D_simul}), we have
proceeded in the following way.

We have constructed a large number of very simple wind-models with variable
\Ha wind emission, in complete analogy to our stationary description 
(Paper~I). The resulting profiles (10 per model) have been analyzed in the
same way as the observed ones, i.e., by means of the TVS-analysis as
described in Sect.~\ref{method}. To allow for a direct comparison with
results from our observations, we have added artificial Gaussian noise to
the synthesized profiles (S/N = 200, which is a typical value), and have
re-sampled the synthetic output onto constant wavelength bins corresponding
to an average resolution of 15\,000.

In order to account for the effects of wind disturbances of different size
and density contrast in different geometries, we calculated various 
series of models: three consisting of spherical shells (series SS) 
and four consisting of broken shells (i.e., clumps, series BS1 and BS2). A
summary of the various models and their designation is given in
Table~\ref{1dmodels}.

All simulations are based on our (quiet) model for HD~188\,209 with $\Mdote
= 1.6 \Mdu$ (cf. Paper I). In order to investigate the reaction of the TVS
of the synthetic profiles as a function of wind-strength we calculated, for
each model series, 9 different models, with $\Delta \log \Mdote \approx
0.1$, particularly at $0.8, 1.25, 1.6, 2.0, 2.5, 3.2, 4.0, 5.0$ and $10.0
\Mdu$. In this way, profile shapes varying from pure absorption via P~Cygni
type to pure emission have been obtained, covering all ``observed'' mean wind
densities.

In all these models, only the density was allowed to be variable, whereas
the velocity field and thus the NLTE departure coefficients (as a function
of velocity) have been kept at their original value. 

In a first step (series SS1/SS2), we used the most simplistic approach of
disturbing the density, namely we varied this quantity in a random way as
a function of radius only, i.e., we assumed spherical shells. This step,
although rather unrealistic, has been performed particularly to
check the stability of results from somewhat more ``sophisticated'' models
(series BS1/BS2), which are described below.

The variations are defined in such a way as to allow for both positive and
negative disturbances around the ``quiet'' model, in order to preserve the
mean profile. We divided the wind into shells of equidistant velocity range,
$\delta v_{shell}$, where 
\beq 
\delta v_{shell} = c \cdot v_{\rm th}{\rm (H)}, \quad c = 0.5,1 
\eeq 
with $v_{\rm th}$(H) the thermal velocity of
hydrogen. The different multipliers $c$ define two different series, SS1 and
SS2, respectively. Inside each of the shells, the density has been perturbed
by a maximum amplitude of $\pm$70\%, 
\beq 
\rho = \rho_0 (1 + \delta\rho/\rho_0), \quad \delta \rho/\rho_0 = 
-0.7 + 1.4 \cdot {\rm RAN}, 
\eeq 
with $\rho_0$ the stationary density and {\sc ran} a random number
uniformly drawn from the interval [0,1]. The specific maximum amplitudes
(for series SS, but also for series BS, see below) have been chosen in such
a way that the resulting TVS-integrals and FEEWs are (roughly) consistent
with the observed values. Lower maximum amplitudes would result in 
a too low degree of variability, and higher ones in too large values.

Note that the density contrast has been assumed to be constant within each
of the shells, i.e., the number of drawn variables is given by the total
number of shells, which is of the order of 80 for \vinf = 1650 km/s and
$c=1$. Note also that only the wind has been allowed to be variable, i.e.,
we considered perturbations only outside the sonic point, located roughly at
20 \kms.

\begin{figure}
\begin{minipage}{7.5cm}
\resizebox{\hsize}{!}
{\includegraphics{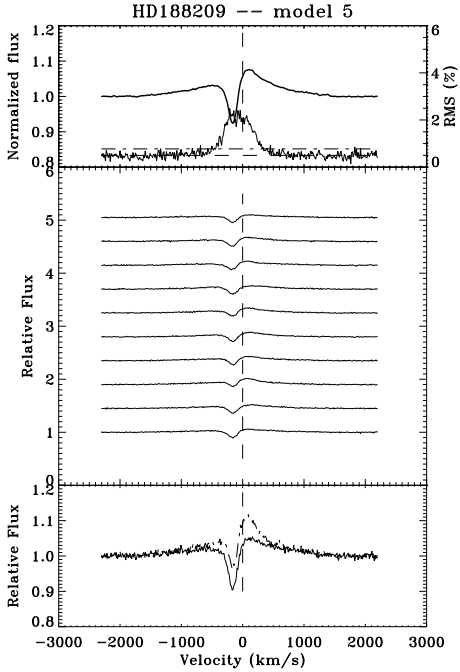}}
\end{minipage}
   \caption{As Fig.~\ref{hots}, but for {\it synthetic} profiles 
   corresponding to model BS11 (broken shells, low amplitude) of
   HD~188\,209, at \Mdot = $2.5 \Mdu$ (see text).} 
\label{TVS_synt} 
\end{figure}

In this way then, different and random amplitudes within the maximum range
$\delta \rho/\rho_0 \in [-0.7,0.7]$ are created for each individual 
shell. To allow for a temporal variability of the resulting profiles at
``random'' observation times (remember that our observations are typically
separated by much more that one wind flow time) we performed, for each model,
10 simulations with different initialization of {\sc ran} and different
locations of the contributing shells. The resulting 10 different profiles
have been analyzed subsequently by means of the TVS method (an example is
given in Fig.~\ref{TVS_synt}).

\begin{figure*}
\begin{minipage}{8.8cm}
\resizebox{\hsize}{!}
   {\includegraphics{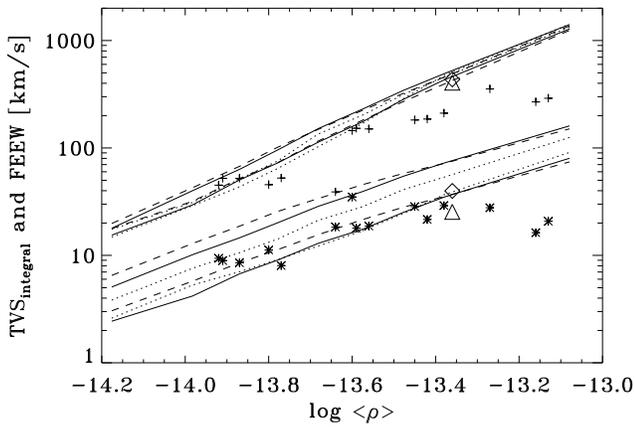}}
\end{minipage}
   \hfill
\begin{minipage}{8.8cm}
\resizebox{\hsize}{!}
   {\includegraphics{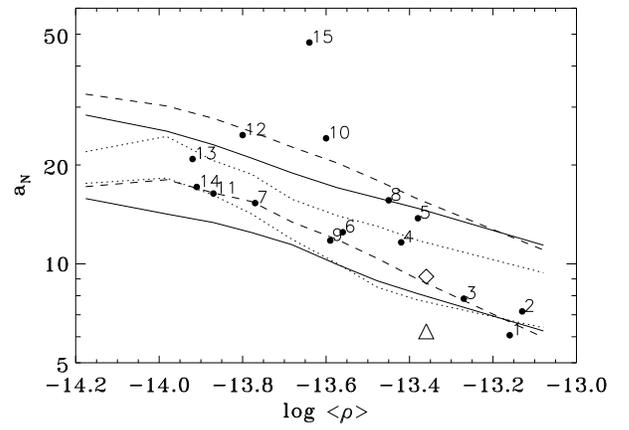}}
\end{minipage}

\caption{{\it Left panel}: ``Observed'' values of the numerator entering
$a_{\rm N}$ (TVS$_{\rm integral}$, asterisks) and of the fractional emission
equivalent width (denominator of $a_{\rm N}$, crosses), compared with
simulated quantities, as a function of mean wind density $\log <\rho>$. The
dotted curves correspond to the simulations with spherical shells and
constant $\delta v$ spacing (SS1/SS2), and the other curves to the somewhat
more realistic simulations accounting for broken shells and constant $\delta
m$, BS11/BS12 (fully drawn) and BS21/BS22 (dashed). Note that inside each
series of simulations the maximum amplitude of density-fluctuations, $\delta
\rho/\rho_0$, is identical, i.e., independent on wind density.
{\it Right panel}: As left, but for the fractional amplitude $a_{\rm N}$ (in
\%). The object numbers correspond to the entries given in Table~\ref{lpv}.
\newline
In both panels, the special symbols correspond to the data resulting from
our TVS-analysis of the 3-D models presented by \citet{Harries}, cf.
Sect.~\ref{3D_simul} (diamonds: spiral structure; triangles: clumpy
structure).} 
\label{comp1_2}
\end{figure*}

Model series SS suffers from (at least) two major problems. At first, the
assumption of {\it spherical} shells might amplify the lpv at all
frequencies, since, especially in the wind lobes, there is only a weak
chance that fluctuations will cancel out due to statistical effects. Second,
our simulations define equal amplitudes of disturbance inside shells of
equidistant range in {\it velocity} space. Accounting for the rather steep
increase in velocity inside and the flat velocity field outside, this means
that the contributing volume per shell is strongly increasing with radius,
which might give too much weight to disturbances in the outer wind.

To ``cure'' both problems, we have calculated four additional model series,
which should be more realistic than the above ones. At first, the spherical
symmetry is broken by the following modification.\footnote{Remember that the
radiative transfer is performed in the usual $p-z$ geometry, with impact
parameter $p$ and height over equator $z$. The so-called {\it core rays} are
defined by $p \le \Rstare$, and the {\it non-core} rays passing {\it both}
hemispheres of the wind lobes by $p > \Rstare$.} For the core-rays, we
assume {\it coherent} shells (blobs), either of a relatively small lateral
extent, $\Delta p \approx\Rstare/10$ (series BS1) or of a larger extent, 
$\Delta p = \Rstare$ (series BS2). For each of the non-core rays
(distributed roughly logarithmically), on the other hand, we assume
different locations of the density variations {\it per ray}, to simulate the
presence of {\it broken} shells. The latter modification results in a lower
TVS particularly in the red part of the profiles, due to cancellation
effects.  Note that we have convinced ourselves that different distributions
of non-core rays gave very similar results.

In order to avoid the volume effect, instead of assuming
$\delta \rho/\rho_0$ as random, however constant per shell of thickness
$\delta v$ = const, we now require that the random perturbations should 
occur in shells of {\it equal mass},
\beq
\delta m_{\rm shell} = 4 \pi r^2 \rho dr,
\eeq
with roughly 50 (broken) shells per model. Inside each broken $\delta m$
shell, the density fluctuations are evaluated as above. For each of our
simulations BS1 and BS2, we have used two different values for the maximum 
amplitude, max$(\delta \rho/\rho_0$) = $\pm 0.35$ and $\pm 0.7$ (BS11/BS21 and
BS12/BS22, respectively), which gives a fair consistency with the range of
observed variability. 

Before we discuss the results in detail, let us already point out here the
major outcome. Although the assumptions inherent to the various model series
(SS vs. BS) are rather different, the results with respect to interesting
quantities are fairly similar. The only difference concerns the distribution
of the variability over the profile. For the spherical shells models, we
find significant variability on the red side, whereas for the broken shell
model the variability extends to larger blue velocities, due to the
increased influence of the shells in front of the disk (cancellation effects
in the lobes, see Fig.~\ref{TVS_synt}). 

In Fig.~\ref{comp1_2} we now compare the outcome of all our simulations
with the observations (``conservative'' values).  On the left panel, we have
plotted the numerator entering $a_{\rm N}$ (lower set of curves), and the
fractional emission equivalent width (FEEW, denominator of $a_{\rm N}$,
upper set of curves) as a function of mean wind density $<\rho>$. Obviously,
series SS and BS give similar results, as already noted. In particular, the
results of models SS1 and models BS11 (lower dotted and fully drawn
curves) are almost identical, which shows that a large number ($\sim
160$) of spherical shells (model SS1) can simulate the outcome of a model 
with a lower number ($\sim 50$) of broken shells and a lower density contrast.
Moreover, it seems that the ``volume effect'' discussed above is
insignificant, simply because \Ha forms in the lower wind region. This
similarity in the results points to a rather large probability that our
results are robust and independent of the specific assumptions.

Interestingly, both the observations (except for the two objects with highest
wind-density and {\it more localized TVS}, HD~190\,429A and HD~16\,691) and
all simulations roughly follow a power-law for both quantities,
\begin{eqnarray}
\log(TVS_{\rm integral}) &\approx& a + b \log <\rho> \nonumber \\
\log({\rm FEEW}) &\approx& c + d \log <\rho> \nonumber,
\end{eqnarray}
which immediately shows that our hypothesis of both quantities being
proportional to each other fails. Note at first that the logarithmic
dependence of the FEEW on $\log <\rho>$ can be readily understood if one
remembers that the {\it total} emission equivalent width of \Ha scales as a
power-law of mean wind-density (cf. Puls et al. 1996), and that the
integration range $[v_b, v_r]$, entering the fractional equivalent width, is
only weakly increasing with $<\rho>$, if scaled to \vinf and evaluated on a
logarithmic scale (see Fig.~\ref{lpv_1}).  Since the TVS and its integral,
on the other side, is also related to the mean wind-density (at least if the
disturbances do not totally decouple from this quantity), the power-law
dependence of this quantity on $<\rho>$ can also be understood. The
different and lower slope can be attributed to optical depth effects and the
cancellation of fluctuations in the emission lobes (vs. the contributions
from core-rays), at least in our simulations (where we ``know'' the origin
of the variability). In total then, $a_{\rm N}$ becomes a decreasing
function of $\log <\rho>$, 
\beq
\log a_{\rm N} \approx (a-c) + (b-d) \log <\rho>,\qquad b < d,
\eeq
and since all our model series predict the same dependency, it is rather
likely that this effect should be present also in more realistic simulations. 
In conclusion, we predict that $a_{\rm N}$ becomes a decreasing function of
mean wind density, {\it even if the disturbances} (more precisely, their
relative amplitudes) {\it are independent of $<\rho>$}. The vertical off-set
of this relation, on the other hand, depends strongly on the density contrast, 
i.e., on max$(\delta \rho/\rho_0)$. All this simulations, of
course, refer to the case of fluctuations which are ``globally'' present,
and will not explain effects from localized macro-structures such as CIRs.

\begin{figure*}
\begin{minipage}{8.8cm}
\resizebox{\hsize}{!}
   {\includegraphics{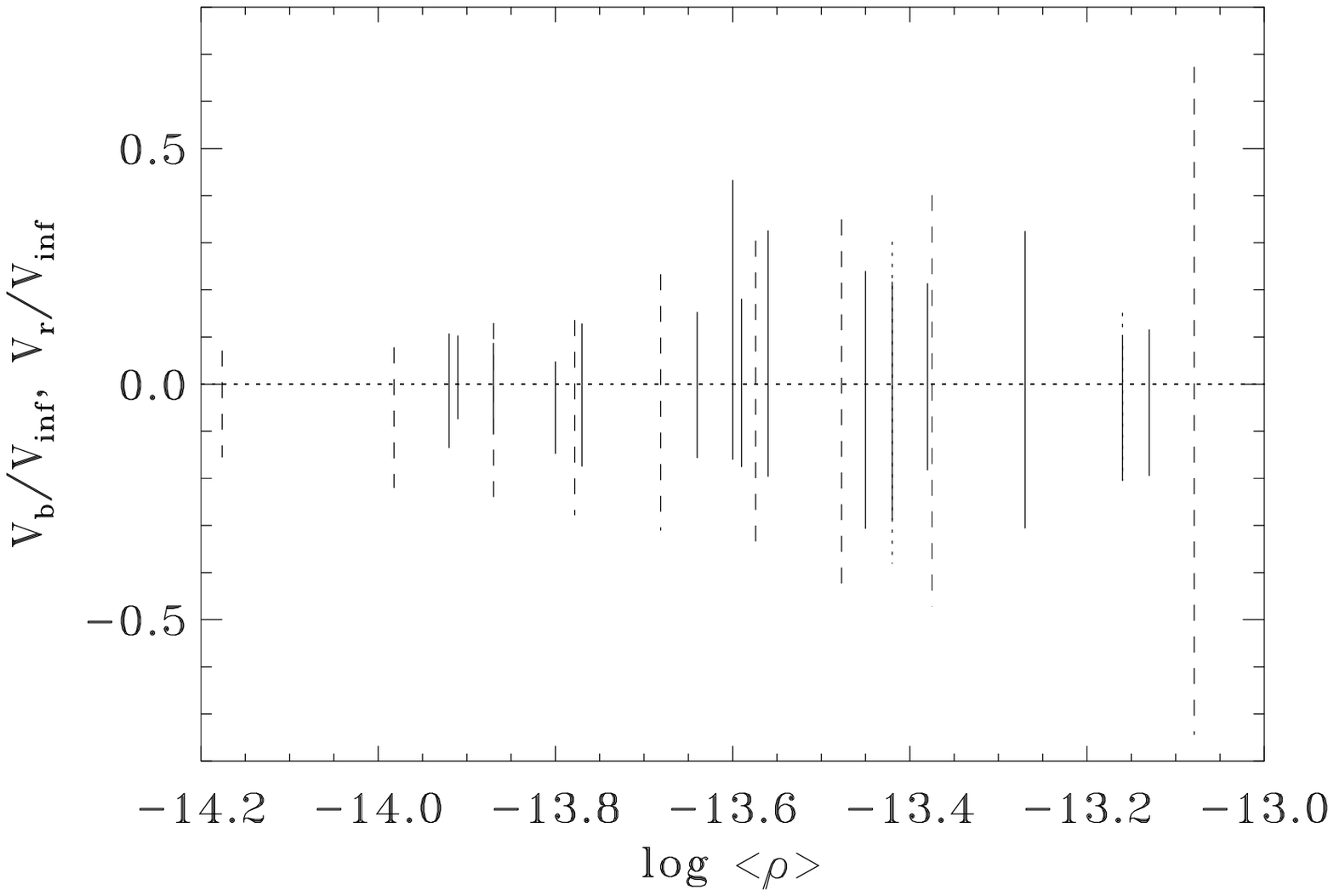}}
\end{minipage}
   \hfill
\begin{minipage}{8.8cm}
\resizebox{\hsize}{!}
   {\includegraphics{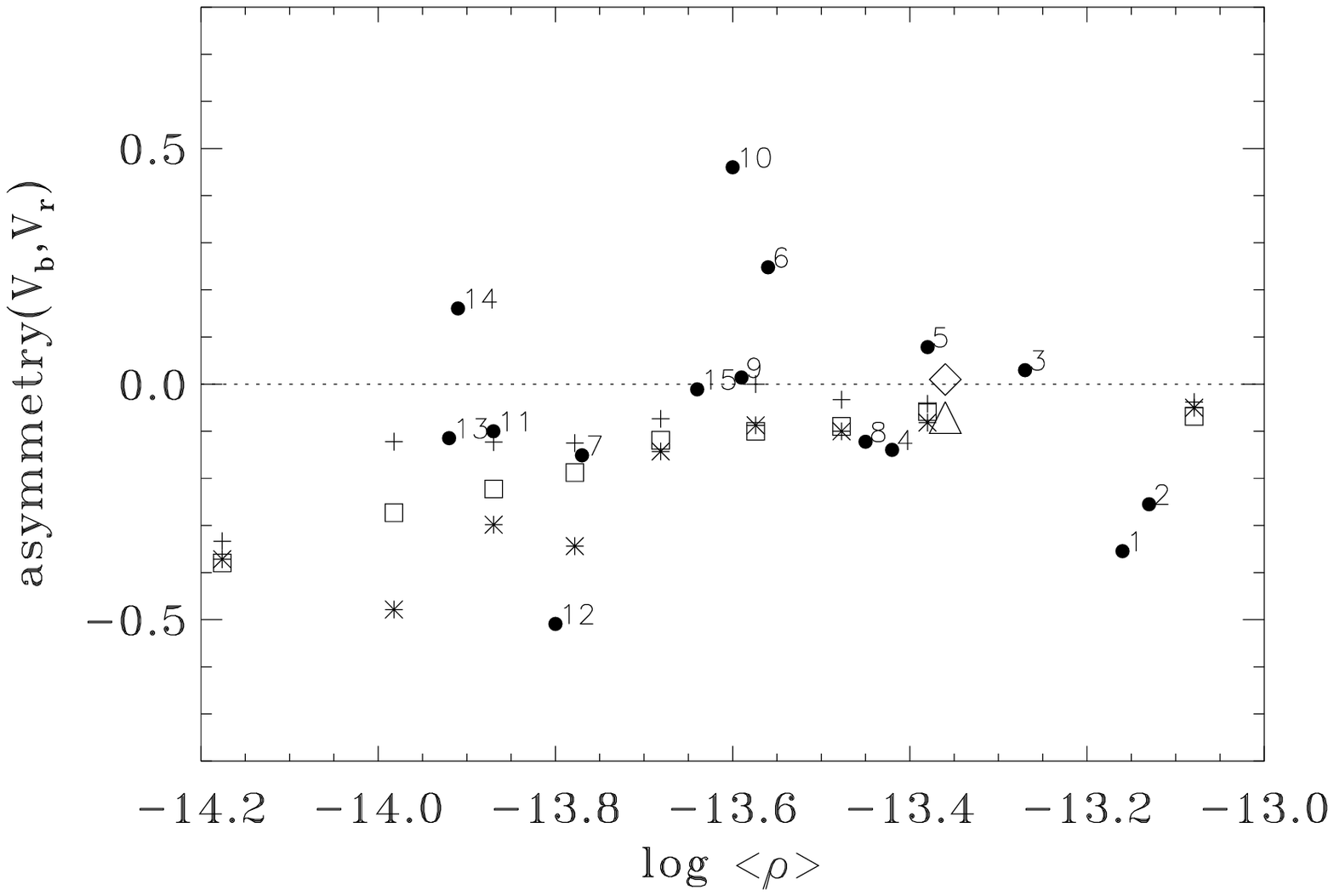}}
\end{minipage}
\caption{{\it Left panel}: Observed blue and red velocity limits, $v_b, v_r$ 
(conservative and non-conservative values) in units of \vinf, as a function
of mean wind-density, compared with results from simulation BS21 (dashed). 
\newline {\it Right panel}: Observed asymmetry, $(v_r-|v_b|)/(v_r +|v_b|)$
(conservative values, black dots), compared with simulations (asterisks:
BS21, squares: BS11, crosses: SS3). The object numbers correspond to the
entries given in Table~\ref{lpv}. The diamond and triangle refer to the data
resulting from our TVS-analysis of the 3-D spiral and clumpy model presented
by \citet{Harries}, cf. Sect.~\ref{3D_simul}.} 
\label{vv_asym}
\end{figure*}

The actual and predicted behaviour of $a_{\rm N}$ is shown in
Fig.~\ref{comp1_2}, right panel. For lower wind densities, the slopes of the
relations for numerator and denominator are rather similar (optically thin
winds, wind emission dominated by core-rays), so that the predicted
amplitude $a_{\rm N}$ remains roughly constant or is even increasing,
whereas from $\log <\rho> \approx -14.0$ the predicted decrease is obvious.
By comparison with observations, we find that almost all stars lie witin the
range suggested by BS11/BS12 and BS21/BS22, i.e., correspond to differences
in (relative) amplitude within a factor of two. There are only two outliers,
HD~210\,809 (\#10) and particularly HD~18\,409 (\#15, at $a_{\rm N} \approx
50$). Remember that the former star has been designated as a ``peculiar''
object in Sect.~\ref{profileshape}, whereas the strong deviation of
HD~18\,409 is more likely due to uncertainties in wind parameters
\citep{Repo04}: the large value of $a_{\rm N}$ is {\it not} due to a large
TVS-integral, but due to a rather small value of its FEEW (located at $\log
<\rho> \approx -13.64$ and FEEW = 39 km/s in the left panel.

Even for the two objects with the largest wind densities, which have been
found not to follow the individual relations for the TVS-integral and the
FEEW (due to their rather localized variability), the results for
$a_{\rm N}$ are consistent with the predictions. In our
interpretation, this would mean that both stars have the same degree of
activity as the other stars, only in different and more localized regions. 

In summary, there are no indications of a dependence of wind activity on wind
density, at least on basis of our present simulations; in our interpretation, 
the decrease of $a_{\rm N}$ is an artefact of the normalization, which 
(unfortunately) does not follow the same slope as the TVS-integral. 

Furthermore, from our simulations, we may also conclude that the underlying
disturbances (at given wind density) may introduce a scatter up to a factor
of two in the maximum amplitude. Of course, more realistic simulations are
needed before a final statement concerning this point can be given. 

Further insight into the origin of the variability might be found from a
comparison of observed vs. predicted velocity limits, $v_b, v_r$ and
particularly of their asymmetry, cf. Fig.~\ref{vv_asym}. In the left panel
of this figure, we compare these velocity limits (normalized to \vinf, to
obtain a unique scale) with our predictions, here with results from BS21
(broken shells, low amplitude). Obviously, for stars with low and 
intermediate wind strength ($[\log <\rho>] < -13.2$) and except for
HD~210\,809 (at $\log <\rho> = -13.6$, with strongly asymmetric velocity
limits), our models do fairly reproduce the observed amount of increase of 
$\Delta V$ as a function of $<\rho>$. Since at largest wind densities we
have only two objects in our sample, it is not clear at present whether their
discrepant behaviour is peculiar or not. Thus, except for the outliers,
it might be concluded that the observed variability results from effects
which are present {\it everywhere in the wind}, in accordance with our
models. This conclusion seems to be also supported by the fact that
HD~210\,809 deviates from our predictions: for this star our observations
have suggested the presence of large-scale wind disturbances which are
localized rather than uniformly distributed over the wind volume. 

The right panel of Fig.~\ref{vv_asym}, however, shows also the shortcomings
of our models. Plotted is the asymmetry of $v_b, v_r$ by means of the
expression $(v_r-|v_b|)/(v_r+|v_b|)$ (negative values correspond to
blue-to-red, positive values to red-to-blue asymmetries, respectively). Let
us firstly discuss the ``theoretical'' predictions. We have plotted the
results for model series BS21 (asterisks), BS11 (squares) and for a model
which has been constructed additionally for this comparison, namely a model
with {\it spherical} shells and $\delta m$ = const(SS3, crosses). All three
models have the same maximum amplitude, max($\delta \rho/\rho_0$) = 
$\pm$0.35. Not surprisingly, the last model shows the least asymmetry,
whereas model BS21 shows the largest one, due to the large lateral extent of
the assumed broken shells in front of the stellar disk. Note that most
models show a blue-to-red asymmetry, even those with spherical shells,
whereas {\it in no case a redwards asymmetry is found}. This predicition, of
course, results from cancellation effects in the lobes, compared to the
situation for core-rays. For large wind-densities, all models converge to
small or even negligible asymmetry, because of the increasing influence of
the lobes. For model SS3 (spherical shells), symmetry is reached earliest,
roughly at $\log <\rho> = -13.5$. 

With respect to the observations, the situation is as follows. At low and 
intermediate wind densities, only four objects display a {\it significant}
asymmetry, mostly to the red, where the highest degree is found for
HD~210\,809 (\#10, see above). The majority of stars, however, show either a
small degree of asymmetry (both to the blue {\it and} to the red) or behave
symmetrically. The predictions of series SS3 are closest to this behaviour.
Let us point out already here that a symmetric TVS at low wind densities can
be alternatively explained by models with co-rotating structures (e.g.,
CIRs), as will be discussed later on. For the two stars with larger wind
densities, the predictions definitely deviate from what has been observed.  

\medskip
\noindent 
In Sect.~\ref{mean_amp} it turned out that the mean amplitude of
deviations, $A_{\rm N}$, seems to be uncorrelated with mean wind density, at
least for our sample. We speculated that if the sources of wind variability
were (uniformly) distributed over the contributing wind volume, one should
see a positive correlation. By means of our simple models, we can
check this conjecture now and may find additional constraints on the origin
of wind variability. 

\begin{figure}
\resizebox{\hsize}{!} 
{\includegraphics{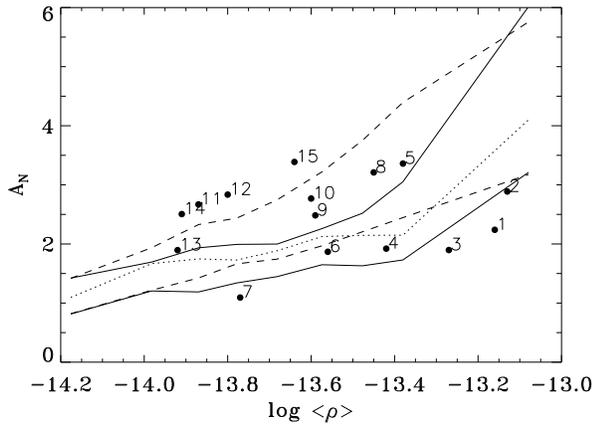}}
\caption{``Observed'' values of the mean amplitude of deviations, $A_{\rm
N}$ (conservative values) compared with simulated quantities, as a
function of mean wind density $\log <\rho>$.  The bold curves correspond to
series BS11/BS12, the dashed curves to series BS21/BS22 and the dotted curve
to series SS3 (spherical shells, low amplitude), respectively. Each series
shows a positive correlation with mean wind density. The
object numbers correspond to the entries given in Table~\ref{lpv}.} 
\label{comp3a}
\end{figure}

\begin{figure*}
\begin{minipage}{8.5cm}
\resizebox{\hsize}{!}
   {\includegraphics{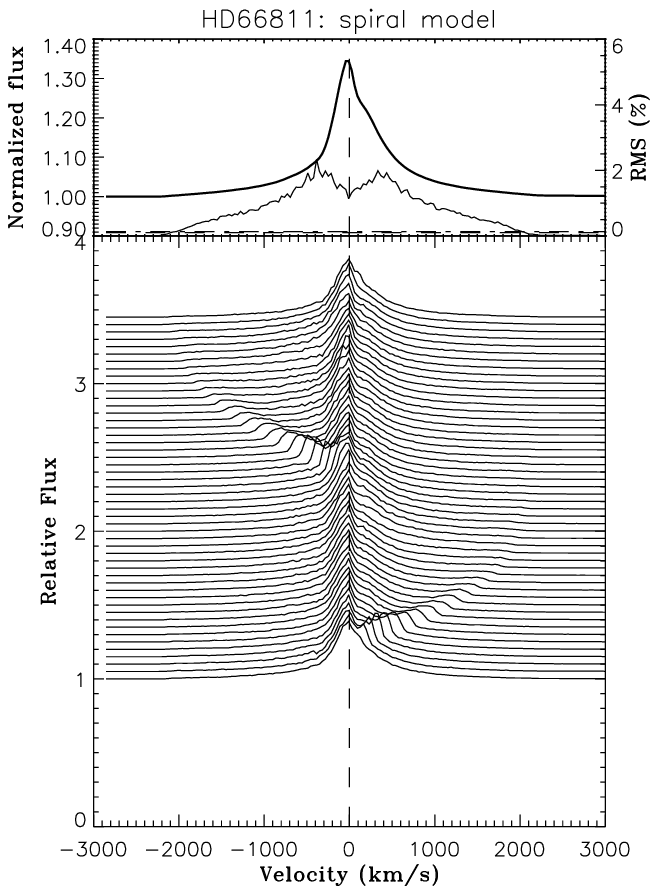}}
\end{minipage}
\hfill
\begin{minipage}{8.5cm}
\resizebox{\hsize}{!}
   {\includegraphics{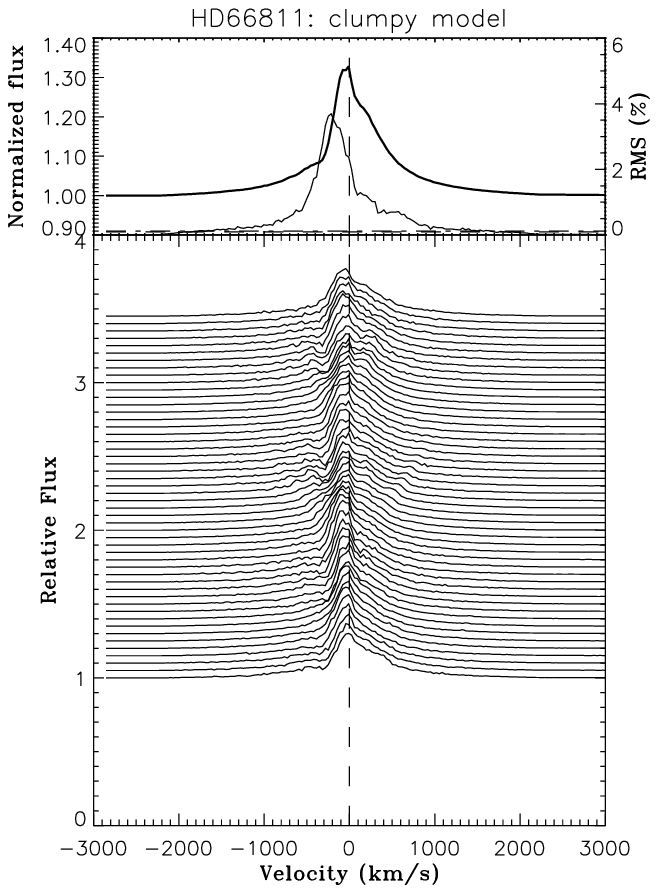}}
\end{minipage}
\caption{As Fig.~\ref{hots}, however for the time-series of synthetic \Ha
profiles from $\zeta$ Pup as calculated by \citet{Harries}, using two
different 3-D models for the wind morphology. {\it Left panel}: wind with
co-rotating, one-armed spiral density enhancement; {\it right panel}:
clumped wind. Note the different distributions of the corresponding TVS.}
\label{3D_simul_plot}
\end{figure*}

Fig.~\ref{comp3a} shows the behaviour of the simulated mean amplitude as a
function of mean wind density, for series BS11/BS12, series BS21/BS22 and SS3,
respectively. Actually, $A_{\rm N}$ {\it is} strongly correlated with
$<\rho>$, where the vertical offset is a function of (relative) amplitude
$\delta \rho/\rho_0$. 
On a first glance, we might conclude that the observed values are again (i.e.,
as we have found for $a_{\rm N}$) consistent with the models, if we allow
for a variation in amplitude of roughly a factor of two. 

A closer inspection, on the other hand, reveals that at least for two low
density objects (HD~188\,209, \#11 and HD~218\,915, \#14) there is the
following problem. The mean amplitudes of these objects are consistent with
our simulations with larger $\delta \rho/\rho_0$ (BS22). Concerning the
fractional amplitudes, $a_{\rm N}$, however, they are consistent with our
simulations for lower $\delta \rho/\rho_0$ (BS21, cf. Fig.~\ref{comp1_2}),
which in turn produce a much too large bluewards asymmetry in the TVS 
(Fig.~\ref{vv_asym}). Remarkably, one of these objects
(HD~218\,915, \#14) even suffers from an observed red-to-blue asymmetry.

\smallskip
\noindent
Thus, from the comparison of mean amplitudes and asymmetry in the
velocity limits we find a number of indications that at least two (from 3) 
low and the high density objects {\it
cannot} be explained by our simple models consisting of {\it density
fluctuations distributed everywhere in the wind}. If we return to our
original TVS analysis (Figs.~\ref{lts} and \ref{hots}), the primary sources
for these inconsistency can be clearly identified: (i) the strong
variability of a central reversal at zero rest velocity, which gives rise
to a rather large TVS within a narrow, symmetric velocity range for stars of
weaker winds and (ii) the rather small extent of the observed variability, 
preferentially bluewards of the emission peak for stars of stronger winds.

In our models, a large variation at rest wavelength cannot be reproduced, at
least if we allow for fluctuations of both positive and negative amplitude.
Such a behaviour might be explained by radially extended, coherent
structures in front of the disk, which would mimic a certain global increase
of mass-loss by bringing the innermost part of \Ha into emission. That our
models can never reproduce a (strongly) localized variability has been
discussed already above.

\subsection{3D model simulations}
\label{3D_simul}

Recently, \citet{Harries} has published results for 3-D line-profile
simulations of \Ha for $\zeta$~Pup, performed by means of his Monte Carlo
stellar wind radiative transfer code. Two distinct models for the wind
morphology have been considered: one with a co-rotating (one-armed) spiral
structure and another one consisting of a clumpy wind. In the first case 
the author examined the effect of one spiral streamline of enhanced density 
on H$\alpha$, while in the latter one he considered random blobs propagating 
throughout the wind. 

In order to obtain an impression to what extent the outcome of our model
analysis might be influenced by the fact that we consider 1-D instead of
more realistic 3-D models, we analyzed the two different sets of synthetic
profiles derived by Harries (kindly made available to us by the author), in
the same way as for the time-series of our sample stars and for our own
simulations. The corresponding results are shown in
Fig.~\ref{3D_simul_plot}.

Apart from the impressive sequence of synthetic profiles which allows to
easily follow any evolution of wind structure in time 
a number of interesting features are noticeable. 

In the first place, the distributions of amplitudes for the two models 
are rather different: in the one-armed spiral model, the derived TVS is 
double-peaked and {\it symmetric with respect to the stellar rest frame}, 
while in the clumped model it is single peaked with maximum amplitudes 
concentrated on the blue. In addition, for the clumped model the velocity 
range of $significant$ variability shows a clear blue-to-red asymmetry
($v_b$=-1670, $v_r$=1440 \kms) while for the spiral model it is almost
symmetrical with respect to the rest wavelength 
($v_b$=-2010, $v_r$=2050 \kms). 

Interestingly, also the parameters derived from the TVS analysis of the
spectra for the 3-D clumped model are quite similar to the ones we have
derived in terms of our 1-D broken shell (i.e., clumpy) models (see
Figures~\ref{comp1_2} and \ref{vv_asym}), whereas the small differences 
might explained by the fact that the spectra from the
3-D simulations are free of noise, in contrast to the spectra from our 1-D
models.

These findings suggest that at least in the case of a clumpy wind the TVS
signatures do not or little depend on the specific model, and that the
results of our simple 1-D simulations are comparable to those from more
realistic simulations. A final conclusion concerning this point can be drawn
only, of course, when a larger set of 3-D simulations for a variety of wind 
parameters, will performed.

From the point of view of the observations, on the other hand, the above
finding means also that the probability to obtain a symmetric TVS of \Ha
will be very low if the wind is clumpy, even if using snapshot observations.

\section{Spectral variability and the Wind Momentum Luminosity
Relationship}
\label{wlr}

The most extreme case of snapshot observations occurs when performing a
spectral analysis, since usually only one spectrum (or two consecutive ones)
is/are available. Because of the spectral variability inherent to the objects
under discussion, these individual data might be not very representative. 
In the following, we shall investigate this point with respect to derived
mass-loss rates, with special emphasis on the question in how far the Wind
Momentum Luminosity Relationship (WLR) of Galactic O-type stars is
influenced.

To this end we assume (i) that the observed variability of \Ha is only due
to processes in the wind (i.e., the contribution of absorption lpv is
neglected) and (ii) that this variability is interpreted in terms of a
variable mass-loss rate. Note that the latter assumption is inherent to a 
typical spectrum analysis, since due to the scarcity of the available
data-set(s) a particular mass-loss rate is derived, which of course might
be not representative. 

Following this philosophy, we derived upper and lower limits of the
mass-loss rate for all our sample stars, from the two most extreme spectra
present in the time-series, and calculated the corresponding limits for the
modified wind-momentum rate. Due to the adopted simplifications (standard
model, no clumping) the obtained values of \Mdot might be overestimated.

The mass-loss rate estimates listed in Table~\ref{lpv} show that the
observed variations in \Mdot range from $\pm$4\% of the mean value, for
stars with stronger winds, to $\pm$16\% for stars with weaker winds. Since
the accuracy of our determinations also depends on the strength of the wind
- $\pm$20\%, for stars with \Ha in emission, and $\pm$30\%, for stars with
pure absorption profiles \citep{Markova04} - we conclude that the observed 
variability of \Mdot does not exceed the errors of individual determinations
and thus is not significant. This result seems somewhat astonishing,
especially in those cases when drastic changes in the \Ha profile shape have
been observed. Note, however, that for not too low wind densities small
changes in \Mdot give rise to large changes both in the profile shape and 
in the EW \citep{Puls96}.

\begin{figure}
\resizebox{\hsize}{!}
{\includegraphics{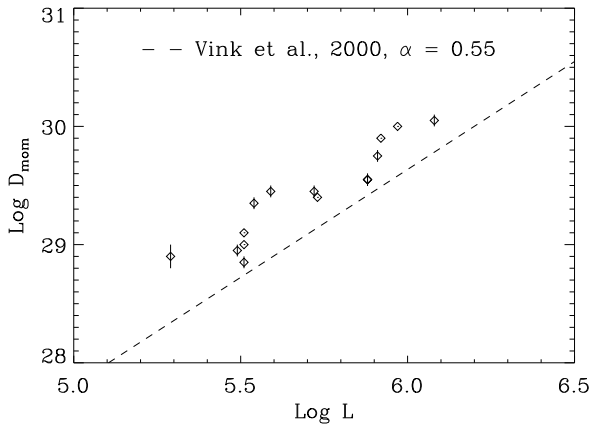}}
\caption{WLR for our sample of Galactic O-type stars. Error bars display
the influence of \Ha lpv (interpreted in terms of \Mdot) on the 
modified wind-momenta.}
\label{wmlr}
\end{figure}

The derived amplitudes of the \Mdot variability have been used to cast
constraints on the variability of the corresponding wind momentum rates. 
The results indicate that wind variability affects the momentum rates by
less than 0.16~dex, which is smaller than the error of individual estimates
(0.30~dex, Markova et al. 2004) and thus is insignificant again. In
addition, and as it can be seen from Figure~\ref{wmlr}, the uncertainty
caused by wind variability does not provide any clue to resolve the problem
concerning the WLR of Galactic O-type stars being a function of luminosity
class \citep{Puls03,Repo04,Markova04}.

Finally, we would like to note that the assumption of a homogeneous and
spherically symmetric wind (underlying our analysis) is in some obvious
contrast with the presence of large-scale perturbations in the winds of some
of our targets, which have been suggested by different investigators on
various occasions (see Sect.~\ref{Ha_time}). Consequently, problems may
occur when trying to fit the \Ha profiles with model calculations.
Exemplary is the case of HD~210\,839, where we failed to
obtain a good fit for a number of profiles of the time series. 

\section{Summary, discussion and conclusions}
\label{summary}

Although line-profile variability of \Ha in O-type supergiants has been
known for a relatively long time (e.g., \citealt{Rosendhal73,Conti77}), our
survey is the first where this variability is detected and quantified in an
objective and statistically rigorous manner for a large sample of stars. By
means of a comparative analysis we furthermore were able to put a number of
constraints on the properties of this variability as a function of stellar
and wind parameters.

The main results of our survey can be summarized as follows:

To study the wind variability in O supergiants, as traced by H$\alpha$, we
employed the TVS analysis, originally developed by \citet{Fullerton96},
however modified in such a way as to account for the effects of wind
emission of various strength on the observed profiles. As already
predicted by these authors, the so-called fractional amplitude of
deviations, which serves as a measure for {\it absorption} lpv, cannot serve
as an indicator for the strength of wind variability. Instead, we
introduced and used a new TVS parameter, $a_{\rm N}$, which measures the
contribution of unit wind emission to the variability detected by the
corresponding TVS. 

By means of this quantity we found that the observed variability of \Ha per
unit wind emission is a moderately decreasing function of mean wind density.
This result might be interpreted as an indication that stronger winds are
less active than weaker ones. However, at least on the basis of our present,
simplistic line-profile simulations, this hypothesis cannot be
supported.

All sample stars show evidence of {\it significant} line-profile variations
in \Ha with {\it mean} amplitudes $A_{\rm N}$ between 1 and 4\% of the
continuum strength. Since absorption lpv seems to be common among O-type
stars \citep{Fullerton96}, these amplitudes can in principle be due to a
combination of variations generated both in the photosphere and in the wind.
However, our analysis indicate that even at lower wind densities 
($\log <\rho> \approx \,-14$) contributions from wind effects have 
to be accounted for, and that for a number of stars wind variability 
is actually the dominating source.

In their original study, Fullerton et al. concluded that the mean amplitude
of deviations is an inappropriate quantity to compare lpv in different lines
of various stars because it does not account for the strength of the
underlying feature. Although true for the case of absorption lpv, this
conclusion has to be somewhat modified with respect to our investigations. 
By means of simple line-profile simulations, the mean amplitude of
deviations in \Ha (defined by Eq.~\ref{A_N}) has been predicted to be an
increasing function of mean wind density, at least in those cases where the
disturbances in density were present everywhere in the wind. Interestingly,
this prediction is {\it not} supported by our observations which gave no
evidence of any correlation between mean amplitude and density. Moreover, it
turned out that a comparison of observed mean and fractional amplitude with
corresponding simulated quantities give different information, because of
the different influence of either the contributing velocity range of {\it
significant} variability or the contributing net wind emission,
respectively. In so far, both quantities (i.e., fractional {\it and} mean
amplitude) deserve their own right and have to be reproduced simultaneously
by future models which claim to explain the observed variability.

\citet{Fullerton96} have found that the {\it distribution} of variability
within an absorption profile can provide information about the cause of the
variations, e.g., radial or non-radial pulsations. By means of a series of 
models with different wind morphology we showed that the TVS analysis of \Ha
can also give some insight into the structure of the wind, at least of its
lower part where this line forms. Both models with spherically symmetric 
and with broken shells produce an \Ha TVS with a blue-to-red asymmetry, if
the structures are distributed uniformly over the contributing wind volume
and as long as the wind-density is not too large. The only
difference in the outcome of the two kinds of models is the actual degree of
the asymmetry: at the same mean density, shell models produce less asymmetry
than those with broken shells (i.e., blobs).  Moreover, this result does not
seem to depend on whether snapshots or systematic simulations have been
used. Note, however, that an asymmetric large-scale, long-lived co-rotating
structure will always produce a symmetric TVS {\it if} the whole period of
rotation is covered by observations. 

A comparison between the observations and line-profile simulations (assuming
variations caused by coherent shells/blobs distributed everywhere in the
wind) revealed that for stars of {\it intermediate} wind density the
parameters of the \Ha TVS are consistent with those from the models if one
allows for a scatter within a factor of two for the maximum (relative)
amplitude of the disturbances at given wind density.

On the other hand, the established disagreement at lower and higher wind
densities might simply point to the presence of wind effects which are not
accounted for in our simulations.  For example, to reproduce the variations
observed in \Ha at lower wind densities one might suggest the presence of a
radially extended, coherent structure in front of the disk which can mimic
variations in the global mass-loss rate. The major problem concerning the
(two!) stars with stronger winds relates to the fact that the observations
yield a TVS which is localized preferentially bluewards of the rest frame
(inside a rather narrow velocity range), whereas our simulations always
produce an almost symmetric TVS instead. This discrepancy might be again
interpreted in terms of a rather confined region of variability close to the
star.

Our analysis shows that {\it significant} variations in \Ha occur below
0.3~\vinf. This estimate is a bit higher than the value derived by
\citet{Kaper97}, namely 0.2~\vinf. Since the velocity 
extent of the observed variations depends strongly on $<\rho>$ and since 
our sample includes a couple of stars with strong winds (\Ha completely in
emission) which are missing in the sample of Kaper et al., such a discrepancy
is quite natural. Moreover, the velocity limits as derived by us, 
converted to units of physical space, correspond to 1.4 to 1.5 \Rstar, in good
agreement with results from theoretical calculations with respect to the
outer limits of \Ha line-formation in O supergiant winds. 

Interpreted in terms of variable mass-loss rates, the observed (partly
extreme) variations in the \Ha wind emission indicate variations in \Mdot of
$\pm$4\% of the mean value for stars with stronger winds and of $\pm$ 16\% 
for stars with weaker winds. The ratio of maximum to minimum 
mass-loss rate determined over the time interval present for our sample ranges
from 1.08 to 1.47, with a tendency that weaker winds show larger values. The
mean ratio averaged over the whole sample is 1.22, in good agreement with the
value provided by \citet{Prinja86}.

The consequences of wind variability with respect to the wind momentum rate
of the sample stars is smaller than 0.16~dex and hence not significant, taken
the individual errors inherent to any \Mdot determination. This result
agrees well with an investigation by \citet{Kudritzki99} who reported
0.15~dex as the error in the wind momentum rate of one(!) A-supergiant,
HD~92\,207, introduced by wind variability. Thus, we conclude that wind
variability in O supergiants does not seem to affect the main concept of 
the WLR although it might contribute to the local scatter by moving
individual points (up and down) with respect to each other.


Finally, it may be worth noting that the \Ha profiles of the sample stars
derived throughout our observations are quite similar, both in shape and
strength, to those obtained by other investigators in various epochs (e.g.,
\citealt{Rosendhal73,Conti77,Scuderi92,Ebbets82, Underhill95,Kaper97}). This
finding suggests that the atmospheres of our targets are not subject to 
long-term variability. On the other hand, no information about previous \Ha
observations concerning the stars BD+56\,739 and HD 338\,926 was found in
the literature.

\acknowledgements
We like to thank the referee. Dr. Alex Fullerton, for his valuable comments 
and suggestions which helped to improve the paper. We are also grateful to 
Dr. T. Harries for providing us with the data from his 3-D wind simulations.
This work was supported in part by a NATO Collaborative Linkage Grant, 
(No. PST/CLG 980007) and by National Scientific Foundation to the Bulgarian 
Ministry of Education and Science (grant No.1407).


\begin{thebibliography}{natbib}
\bibitem[Bianchi \& Garcia(2002)]{BG02} %
        Bianchi, L. \& Garcia, M. 2002, ApJ 581, 610 
\bibitem[Conti \& Frost(1977)]{Conti77} %
         Conti, P.S., \& Frost, S.A. 1977, ApJ 212, 728 
\bibitem[Crowther et al.(2002)]{Crowther02} %
        Crowther, P.A., Hillier, D.J., Evans, C.J., et al. 2002, ApJ 579, 774
\bibitem[Dessart \& Owocki(2003)]{Dessart03} %
        Dessart, L., Owocki, S.P. 2003, A\&A 406, L1
\bibitem[Ebbets(1982)]{Ebbets82} %
         Ebbets, D. 1982, ApJS 48, 399
\bibitem[Feldmeier(1995)]{Feldm95} %
        Feldmeier, A. 1995, A\&A 299, 523
\bibitem[Fullerton et al.(1992)]{Fullerton92}
	Fullerton,A.,W., Gies, D.R. \& Bolton, C.T. 1992, ApJ, 390, 650
\bibitem[Fullerton, Gies \& Bolton(1996)]{Fullerton96} %
        Fullerton, A. W., Gies, D.R \& Bolton, C.T. 1996, ApJS 103, 475
\bibitem[Gies(1987)]{Gies87} %
        Gies, D.R. 1987, ApJS 64, 545
\bibitem[Groenewegen \& Lamers(1989)]{Groene89} %
        Groenewegen, M.A.T., Lamers, H.J.G.L.M. 1989, A\&AS 79, 359
\bibitem[Hamann(1980)]{Hamann80} %
        Hamann, W.-R. 1980, A\&A 84, 342
\bibitem[Harries(2000)]{Harries} %
        Harries, T. 2000, MNRAS 315, 722
\bibitem[Herrero et al.(1992)]{Herrero92} %
        Herrero, A., Kudritzki, R.-P., Vilchez, J.M., et al. 1992, A\&A 261, 209
\bibitem[Israelian et al.(2000)]{Israelian00}    %
        Israelian, G., Herrero, A., Musaev, F., et al. 2000, MNRAS 316, 407.
\bibitem[de Jong et al.(1999)]{deJong99}    %
	de Jong, J.A., Henrichs, H.F., Schrijvers, C. et al., 1999, A\&A 345, 172
\bibitem[de Jong et al.(2001)]{deJong01} %
	de Jong, J.A., Henrichs, H.F., Kaper, L., et al. 2001, A\&A 368, 601
\bibitem[Kaper et al.(1997)]{Kaper97} %
	Kaper, L., Henrichs, H.F., Fullerton, A.W., et al. 1997, A\&A 327, 281
\bibitem[Kaper et al.(1999)]{Kaper99} %
        Kaper, L., Henrichs, H.F., Nichols, J.S. \& Telting, J.H. 1999 A\&A, 344, 231
\bibitem[Kaufer et al.(1996)]{Kaufer96} %
        Kaufer, A., Stahl, O., Wolf, B., et al. 1996, A\&A 305,887
\bibitem[Kudritzki(1999)]{Kudritzki99} %
        Kudritzki, R.-P. 1999, in Proc. IAU Coll. No. 169, eds. B. Wolf, O. Stahl 
	\& A.W. Fullerton,  Sprimger Verlag, p. 405
\bibitem[Kudritzki \& Puls(2000)]{KP2000} %
        Kudritzki, R.-P., Puls, J. 2000, ARA\&A 38, 613
\bibitem[Markova(1986)]{Markova86} %
        Markova, N. 1986, A\&A 162, L3
\bibitem[Markova(2002)]{Markova02} %
        Markova, N. 2002, A\&A 385, 479
\bibitem[Markova \& Valchev(2000)]{MV} %
         Markova, N. \& Valchev, T. 2000, A\&A 363, 995
\bibitem[Markova et al.(2004)]{Markova04} %
        Markova, N., Puls, J., Repolust, T., et al. 2004, A\&A 413, 693
\bibitem[Massa et al.(1995)]{Massa95} %
         Massa, D., Fullerton, A.W., Nichols, J.S. et al. 1995, ApJ 452, L53
\bibitem[Meynet et al.(1994)]{Meynet94} %
         Meynet, G., Maeder, A., Schaller, G. et al. 1994, A\&AS 103, 97
\bibitem[Owocki \& Puls(1999)]{OP99} %
        Owocki, S.P. \& Puls, J. 1999, ApJ 510, 355
\bibitem[Owocki, Castor \& Rybicki(1988)]{O88} %
        Owocki, S.P., Castor,J.I. \& Rybicki, G.B. 1988, ApJ 335, 914
\bibitem[Press et al.(1992)]{Press92} 
        Press, W.H., Flannery, B.P.,Teukolsky, S.A. et al. 1992,
	Numerical Recipes: The Art of Scientific Computing (2nd edition),
	Cambridge: Cambridge Univ. Press, p. 634f.
\bibitem[Prinja \& Howarth(1986)]{Prinja86} %
        Prinja, R.K. \& Howarth, I.D. 1986, ApJS 61, 357 
\bibitem[Prinja \& Smith(1992)]{PS92} %
        Prinja, R.K. \& Smith,L.J. 1992, A\&A 266, 377
\bibitem[Prinja \& Fullerton(1994)]{Prinja94}	
	Prinja, R.K. \& Fullerton, A. 1994, ApJ, 426, 345
\bibitem[Prinja et al.(1992)]{Prinja92} %
        Prinja, R.K., Balona,L.A., Bolton, C.T. et al. 1992, ApJ 390, 266
\bibitem[Prinja et al.(1996)]{Prinja96} %
        Prinja, R.K., Fullerton, A.W. \& Crowther, P.A. 1996, A\&A 311, 264
\bibitem[Prinja et al.(2001)]{Prinja01} %
        Prinja, R.K., Stahl, O., Kaufer, A. et al. 2001, A\&A 367, 891
\bibitem[Prinja et al.(2002)]{Prinja02} %
        Prinja, R.K., Massa, D. \& Fullerton, A.W. 2002, A\&A 388, 587
\bibitem[Puls et al.(1996)]{Puls96} %
          Puls, J., Kudritzki, R.-P., Herrero, A., et al. 1996, A\&A 305, 171
\bibitem[Puls et al.(2003)]{Puls03} %
        Puls, J., Repolust, T., Hoffmann, T., et al. 2002, in
	Proc. IAU Symp. No. 212, eds. K.A. van der Hucht, A. Herrero \&
	C. Esteban, ASP Conf. Ser, p. 61
\bibitem[Rauw et al.(2001)]{Rauw01} %
	Rauw, G. , Morrison, N.D., Vreux, J.-M. et al. 2001, A\&A 366, 585 
\bibitem[Repolust et al.(2004)]{Repo04} %
        Repolust, T., Puls, J., Herrero, A. 2004, A\&A 415, 349
\bibitem[Rosendhal(1973)]{Rosendhal73} %
        Rosendhal, J.F. 1973, ApJ 186, 909
\bibitem[Scuderi et al.(1992)]{Scuderi92}
	Scuderi, S., Bonanno, G. , Di Benedetto, R. et al. 1992, ApJ 392, 201
\bibitem[Underhill(1995)]{Underhill95} %
        Underhill, A.B., 1995, ApJS 100, 433	
\bibitem[Vink et al.(2000)] {Vink00}%
        Vink, J.S., de Koter, A., Lamers, H.J.G.L.M. 2000, A\&A 362, 295
\bibitem[Walborn(1973)]{Walborn73}
        Walborn, N. R. 1973, AJ, 78. 1067
\end{thebibliography}
\end{document}